\documentclass[%
 reprint,
 amsmath,amssymb,
 aps,
]{revtex4-2}
\usepackage{enumitem}
\usepackage{graphicx}
\usepackage{dcolumn}
\usepackage{bm}

\usepackage{multirow}

\begin{document}

\preprint{APS/123-QED}

\title{Revealing the core-periphery structure of cities}

\author{Federica Fanelli}
 \affiliation{Sapienza Univ. of Rome, Physics Dept, Piazzale A. Moro, 2, 00185, Rome Italy \\ Sony Computer Science Laboratories - Rome, Joint Initiative CREF-Sony, Centro Ricerche Enrico Fermi, Via Panisperna 89/A, 00184, Rome, Italy \\ 
Centro Ricerche Enrico Fermi (CREF), Via Panisperna 89/A, 00184, Rome, Italy}
\author{Hygor P. M. Melo}%
 %\email{Second.Author@institution.edu}
\affiliation{Sony Computer Science Laboratories - Rome, Joint Initiative CREF-Sony, Centro Ricerche Enrico Fermi, Via Panisperna 89/A, 00184, Rome, Italy \\
Centro Ricerche Enrico Fermi (CREF), Via Panisperna 89/A, 00184, Rome, Italy \\
Postgraduate Program in Applied Informatics, University of Fortaleza, 60811-905, Fortaleza, CE, Brazil}
\author{Matteo Bruno}
\affiliation{Sony Computer Science Laboratories - Rome, Joint Initiative CREF-Sony, Centro Ricerche Enrico Fermi, Via Panisperna 89/A, 00184, Rome, Italy \\
Centro Ricerche Enrico Fermi (CREF), Via Panisperna 89/A, 00184, Rome, Italy}
\author{Vittorio Loreto}
\affiliation{Sapienza Univ. of Rome, Physics Dept, Piazzale A. Moro, 2, 00185, Rome Italy \\
Sony Computer Science Laboratories - Rome, Joint Initiative CREF-Sony, Centro Ricerche Enrico Fermi, Via Panisperna 89/A, 00184, Rome, Italy \\
Centro Ricerche Enrico Fermi (CREF), Via Panisperna 89/A, 00184, Rome, Italy \\
Complexity Science Hub, Josefst\"{a}dter Strasse 39, A 1080, Vienna, Austria}

\date{\today}

\begin{abstract}
The distribution of urban services reveals critical patterns of human activity and accessibility. Proximity to amenities like restaurants, banks, and hospitals can reduce access barriers, but these services are often unevenly distributed, exacerbating spatial inequalities and socioeconomic disparities. In this study, we present a novel accessibility measure based on the spatial distribution of Points of Interest (POIs) within cities. Using the radial distribution function from statistical physics, we analyze the dispersion of services across different urban zones, combining local and remote access to services. This approach allows us to identify a city's central core, intermediate areas or secondary cores, and its periphery. Comparing the areas that we find with the resident population distribution highlights clusters of urban services and helps uncover disparities in access to opportunities.
\end{abstract}

%\keywords{Suggested keywords}%Use showkeys class option if keyword
                              %display desired
\maketitle

\section{\label{sec:introduction}Introduction}

Defining precise boundaries of a city is crucial for establishing metrics, rankings, and comparing statistical data accurately. Likewise, understanding the geographical boundaries of different urban zones—such as central areas, neighborhoods, and outskirts—is complex. Cities are dynamic, constantly evolving, and the focal points of urban activity shift in intricate ways over time.

The definition of metropolitan areas often hinges on identifying their centers, which are typically linked to areas with the highest human activity, though these may not align with the geographical center. Some cities present a monocentric structure, concentrating most activities in a single area, while others display a polycentric structure with multiple centers shaping the urban landscape~\cite{parr2004polycentric}. Several models have been proposed to quantify this phenomenon~\cite{mcmillen2003number, louail2014mobile, odland1978conditions}, including models based on economic interactions between households and firms~\cite{fujita1982multiple}, the correlated percolation model to examine urban interconnectedness~\cite{makse1995modelling}, and traffic congestion as a driving factor in the transition from monocentric to polycentric cities~\cite{louf2013modeling, louf2014congestion}.

Understanding urban structures is crucial in economics, as it sheds light on regional development dynamics, including the formation of core-periphery patterns~\cite{krugman1991increasing, krugman1992geography, bettencourt2021introduction}. The spatial distribution of Points of Interest (POIs) within cities is key to identifying their overall layout~\cite{deng2019detecting}. Moreover, a city's functionality is closely tied to how people move within it, with mobility playing a critical role in defining urban areas. Recently, analyses of citizen movement have been enhanced by data sources such as cell phone records~\cite{alhazzani2021urban, toole2012inferring}, metro and bus card data~\cite{long2015discovering, roth2011structure}, and taxi trip data~\cite{pan2012land}.

City centers can often be identified through accessibility, defined by proximity to urban services. Several measures of location-based accessibility exist, including the spatial distribution of activities~\cite{hansen1959accessibility}, transportation costs~\cite{vickerman1974accessibility}, combined costs and opportunities~\cite{wu2020unifying}, spatial distances~\cite{ingram1971concept}, time in the context of chrono-urbanism and the 15-minute city paradigm~\cite{abbiasov202215,Bruno_et_al_2024}, and public transportation travel times~\cite{biazzo2019general}. Inequality in accessibility~\cite{Bruno_et_al_2024} is also critical for evaluating economic and social disparities. Analyzing how different groups within a city can access critical resources and opportunities, we can uncover patterns of inequality that might otherwise remain concealed~\cite{biazzo2019general, weiss2018global, vale2023accessibility}.

Our study utilizes the radial distribution function~\cite{hansen2006theory, tuckerman2010statistical}, a fundamental concept in physics, which has been applied across various fields~\cite{lin2004k, srolovitz1981radial, kirkwood1935statistical}. We adapt this framework to analyze the radial distribution of Points of Interest across different urban locations. By examining radial distributions from multiple starting points within a city, we can detect the clustering of activities and derive a quantitative measure of accessibility at each point. We then integrate these metrics with population data to assess the inequality of accessibility within cities. Additionally, we employed a method that combines the definition of our metric with a clustering technique to identify distinct urban components within the city structure, and thus its core(s) and periphery.

The main objective of this paper is to introduce a new method for identifying a city's functional core(s) by observing the spatial distribution of POIs. We introduce a novel accessibility measure founded upon the observation of clustering, and we leverage it to propose a method to unveil the city structure, distinguishing a main (or functional) core, intermediate areas or secondary cores, and the periphery of a city.

\section{\label{sec:methods}Methods}
We conducted an analysis of ten cities (Rome, Paris, Barcelona, Atlanta, Fortaleza, Milwaukee, Naples, Bergamo, Vienna, and Bogota) introducing a novel location-based accessibility definition and a method to identify spatial patterns in the distribution of Points of Interest (POIs). A full description of the data employed in this analysis is available in the Appendix.

\subsection{\label{sec:rdf}Radial distribution function}

The main analysis tool used is the radial distribution function (RDF or $g(r)$)~\cite{hansen2006theory, tuckerman2010statistical}, which is a fundamental concept in physics. The RDF is widely used across various fields~\cite{lin2004k, srolovitz1981radial, kirkwood1935statistical}, and it is a versatile tool for examining the spatial arrangement of points.

In order to analyze urban areas, we can consider a city as a specific arrangement of Points of Interest spread throughout the city. Each POI represents a point in space, defined by its latitude and longitude coordinates. These POIs are distributed in a way that reflects the city's characteristics.

\begin{figure*}
    \centering

    \includegraphics[width=\textwidth]{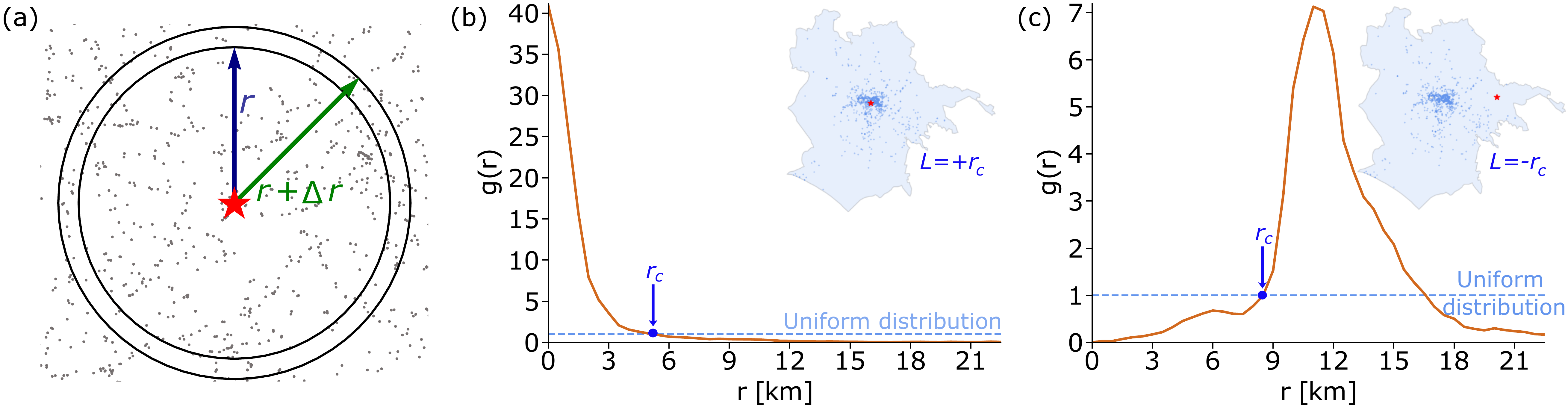}
    \caption{\textbf{An example of the radial distribution function employed to measure accessibility.} (a): graphical representation of the radial distribution function’s idea. The grey dots represent the distribution of POIs in the city, the red star corresponds to the initial point, the shell is depicted with the two concentric black circles and the blue and green vectors are respectively the distance between the starting point and the first and the second circle of the shell. The density of points in the shell normalized by the total density is the measure $g(r)$ of the radial function. (b) and (c): radial distribution function for a point in the city center (b) and a point near the boundary (c) in Rome, with the indication of the $L$ value. The map represents the POIs for the selected category ("Accommodation") and the red star on it indicates the starting point.}
    \label{fig 1}
\end{figure*}

The radial distribution function at a distance $r$ from a fixed point $i$ can be determined by counting the number of points within a shell of radius $r$ and thickness $\Delta r$ from $i$. This can be visualized in Fig.~\ref{fig 1}(a). We calculate the radial distribution of POIs from a point located at the center of cell $i$ in the city grid for all distances $r$, using a fixed value for $\Delta r$.

The radial distribution function $g(r)$, for a fixed starting point $i$ and for a shell centered on $i$ of radius $r$ and thickness $\Delta r$, is:
\begin{equation}
    g(r)=\frac{\rho(r)}{\rho_\text{uniform}},
    \label{g r our}
\end{equation}
where $\rho_\text{uniform}$ is the density of POIs if they were distributed uniformly within the city boundaries. It can be calculated as the ratio between the total number $N_{tot}$ of POIs in the city divided by the area $A_{tot}$ of the city:
\begin{equation}
    \rho_\text{uniform}=\frac{N_{tot}}{A_{tot}},
\end{equation}
and $\rho(r)$ is the real density of POIs in the shell of radius $r$ and thickness $\Delta r$. It is given by the number $N_{\Delta r}(r)$ of POIs counted in this shell divided by the area $A_{\Delta r}(r)$ of the shell:
\begin{equation}
    \rho(r)=\frac{{N_{\Delta r}(r)}}{A_{\Delta r}(r)}.
\end{equation}

The distribution is calculated by gradually increasing the value of $r$ by $\Delta r$ up to the value corresponding to the largest distance between the center of the hexagons and the POIs. The function $g(r)$ has been computed for the center of each hexagon. Two examples of $g(r)$ calculated for different starting points can be found in Fig.~\ref{fig 1}(b) and ~\ref{fig 1}(c).

When moving from a central point to a point at the periphery, we can expect a variation in the radial distribution. It's important to quantify these differences in terms of RDF to describe the city in terms of accessibility.

We approximate the urban area as a 2-D surface and apply border corrections to calculate $A_{\Delta r}(r)$ to account for the irregular surfaces given by urban boundaries.

We conducted a preliminary analysis for each city to choose the optimal value of $\Delta r$ as a compromise between good resolution (small fluctuations) and no loss of information. Please refer to the Supplemental Material~\cite{SM} for a thorough analysis of our methodology.

To minimize fluctuations, we employed the sliding window method to average the values of $g(r)$. This involved taking five consecutive values of $g(r)$, calculating their average, and assigning this value to the corresponding $r$ of the first value within the selected range.

Due to this correction, in some cases, the $g(r)$ curve lies entirely above one or completely below one. In such cases, we take $L$ as the value of $r$ corresponding to the $g(r)$ value closest to one.

\subsection{\label{sec:L def}L-values definition}
Following the interpretation of the radial distribution function $g(r)$ in Eq.~\ref{g r our} as a comparison between the real distribution of POIs in the city and the random uniform distribution of POIs in the same area, we can identify three cases:

\begin{itemize}
    \item $g(r)>1$: the actual density $\rho(r)$ of POIs is higher than the uniform density $\rho_\text{uniform}$, indicating that at distance $r$, POIs are more likely to be found compared to what would be expected in an uncorrelated random arrangement.
    \item $g(r)\approx 1$: the actual density $\rho(r)$ of POIs is equal to the uniform density $\rho_\text{uniform}$, suggesting that POIs are approximately uniformly distributed.
    \item $g(r)<1$: the actual density $\rho(r)$ of POIs is smaller than the uniform density $\rho_\text{uniform}$, indicating a depletion zone.
\end{itemize}

When analyzing the radial distribution functions calculated from different starting points within the cities, two main patterns in the curves can be identified. Some points present a curve that starts with values greater than one and decreases, reaching one at a certain point, as seen in Fig.~\ref{fig 1}(b), while other points show a rising curve from values less than one to values greater than one, as shown in Fig. \ref{fig 1}(c). 

We define $r_c$ as the radius $r$ where the RDF crosses one for the first time:
\begin{equation}
    r_c=\min \{r | g(r) = 1\}.
\end{equation}

As $r$ takes discrete values, an interpolation method was used to find the nearest \( r \) value where \( g(r) = 1 \).

If the $g(r)$ curve is greater than one and decreasing for low $r$, $r_c$ represents the distance at which the surrounding clustering disappears into depletion. If instead the curve is lower than one and increasing for low $r$, $r_c$ represents the distance at which clustering first emerges. To differentiate between the two cases, we define a measure $L$ as follows:

\begin{equation}
    L=
    \begin{cases}
        r_c \ \ \ \ \ \ \ &\text{if } g(r) > 1 \ \ \forall r < r_c; \\
        -r_c \ \ \ \ \ &\text{if } g(r) < 1 \ \ \forall r < r_c.
    \end{cases}
\end{equation}

Two examples of the definition of $L$ can be found in Fig.~\ref{fig 1}(b) and~\ref{fig 1}(c).

We can interpret the parameter $L$ as a metric of accessibility. Positive values of $L$ indicate clustering near the origin point, suggesting a higher probability of finding a POI within a certain distance $r < L$ compared to a uniform distribution. Therefore, positive $L$ values indicate favorable access to POIs, with large values indicating a large area of clustering of POIs around the starting point. On the other hand, negative $L$ values suggest the existence of a depletion zone within a distance of $|L|$, indicating poorer access to POIs and the need to move to find a center of activity.
 
\begin{figure*}%[ht]
    \centering
    \includegraphics[width=\textwidth]{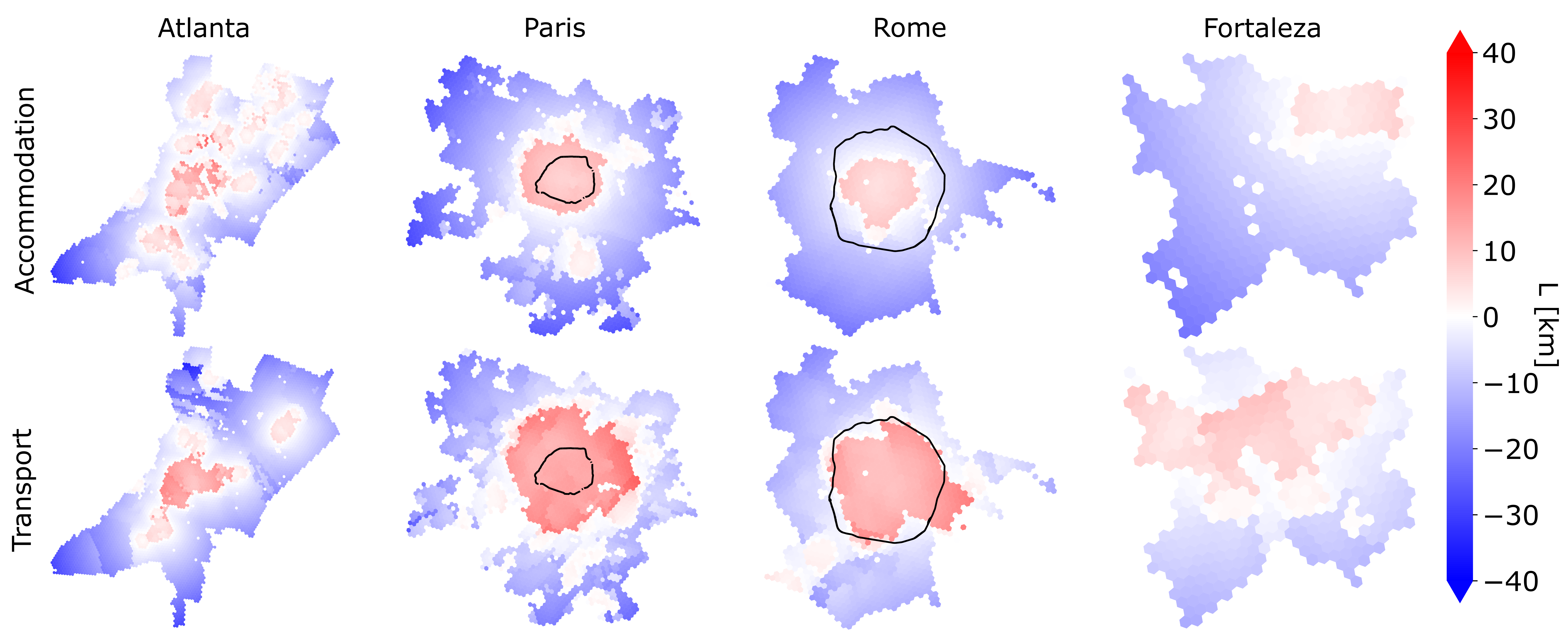}
    \caption{The maps display $L$ values for two categories: "Accommodation" and "Transport". The first column shows data for Atlanta, the second for Paris, the third for Rome, and the fourth for Fortaleza. Additionally, there are black lines on the Rome maps, indicating the Grande Raccordo Anulare (a ring road encircling Rome), and on the Paris maps, representing the Boulevard Périphérique (a similar ring road surrounding Paris). These lines serve as visual references for the inner parts of the two cities.}
    \label{fig 2}
\end{figure*}

 \begin{figure*}
    \centering
    \includegraphics[width=\textwidth]{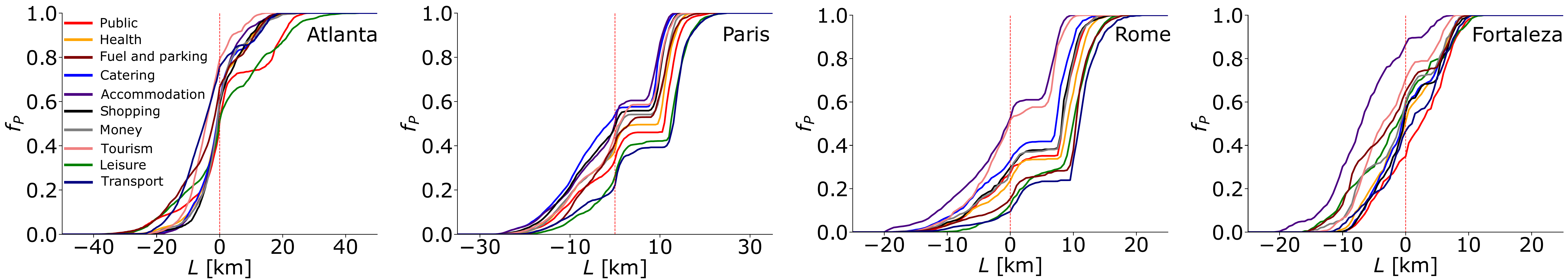}
    \caption{The graph shows the cumulative distribution of population ($f_P$) for different values of $L$ in Atlanta, Paris, Rome, and Fortaleza. The curves reach a value of one at different points on the $L$ axis. Also, the curves for categories related to essential services ("Transport", "Fuel and parking", etc.) intersect the $L=0$ point at lower values of $f_P$ compared to categories related to other services ("Tourism", "Accommodation", etc.). This means that a lower fraction of the population ($f_P$) lives in areas with negative $L$ when considering essential services.}
    \label{fig 3}
\end{figure*}

\section{\label{sec:results}Results}

\subsection{\label{sec:accessibility maps}Accessibility values}

Maps displaying the values of $L$ across cities can be seen in Fig.~\ref{fig 2}, comparing the categories of \textit{Accommodation} and \textit{Transport}. The maps of all cities and categories considered can be found in the Supplemental Material~\cite{SM}. The maps indicate that positive values of $L$ are primarily observed in city centers and surrounding areas, as well as in other disconnected centers such as the maritime district of Rome, Ostia, or the surrounding centers of the Atlanta functional urban area. On the other hand, negative values of $L$ are characteristic of peripheral areas. It is also evident that essential services and facilities necessary for urban living have broader accessibility ranges and higher values of $L$ compared to categories related to consumer services and experiences. It's important to note that our methodology does not assign the highest $L$ values to the city centers, but rather to the border areas of the largest centers, where the metric presents a discontinuity. This spatial discontinuity in the metric indicates the end of a POI clustering area and the beginning of the depletion zone, which we will utilize to identify city cores in section~\ref{sec:cores}.
\begin{figure*}
    \centering
\includegraphics[width=\textwidth]{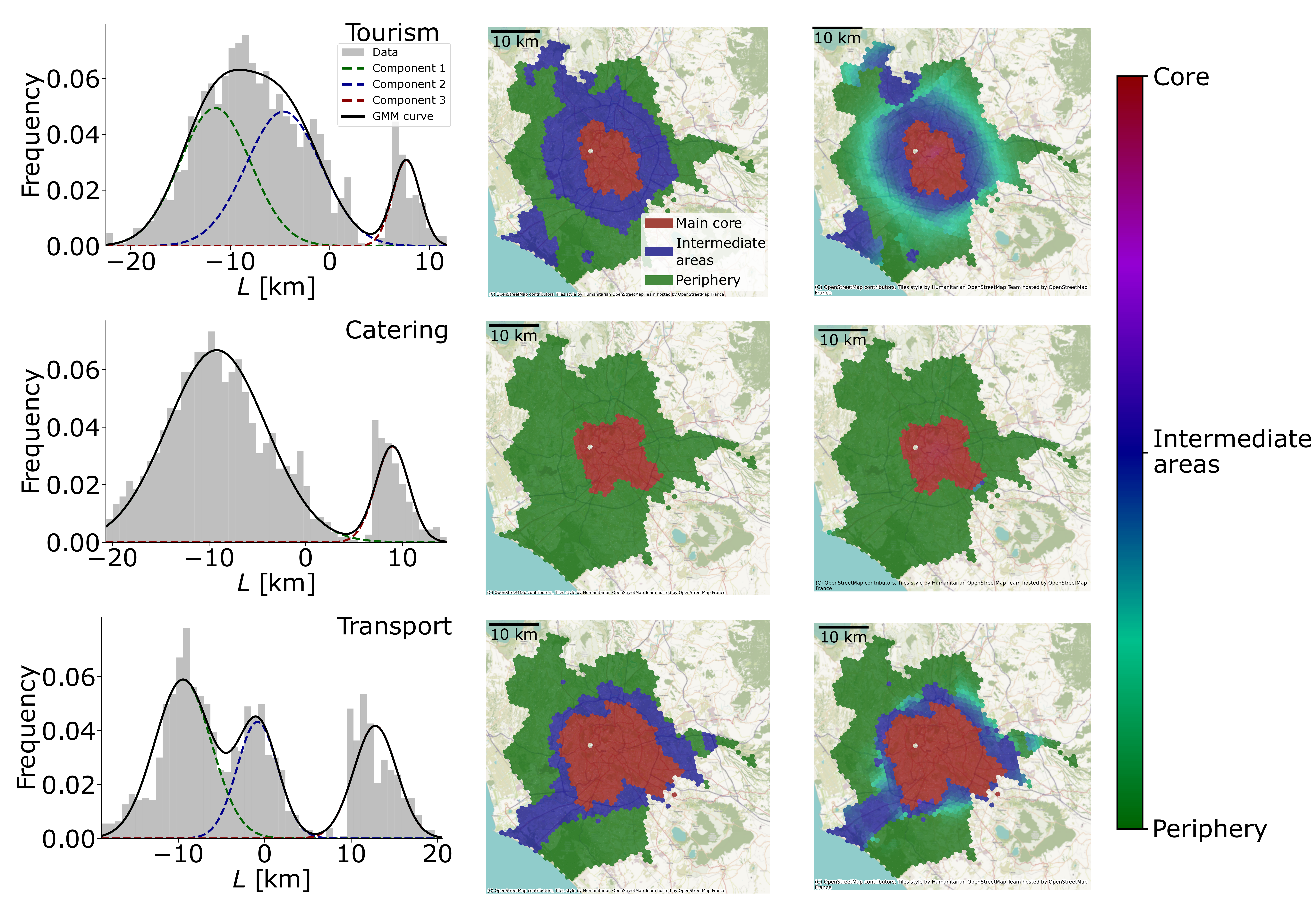}
    \caption{\textbf{Distribution of L values and maps of core-periphery areas in Rome.} On the left, histograms of $L$ values for different categories in Rome. These histograms include a final Gaussian mixture model curve and its Gaussian components.  In the middle, there are maps of Rome with the spatial grid colored according to the method used to identify the components of the urban structure. Hexagons in red represent the main core, identified as the cluster with the highest probability of belonging to the Gaussian component with the highest mean. Areas shaded in green indicate the periphery, characterized by the highest probability of belonging to the Gaussian component with the lowest mean. Blue hexagons represent intermediate areas associated with the Gaussian component in the middle when it is present. In all cases, the main core corresponds to the city center.  On the right, the maps of Rome show the spatial distribution of urban zones, using a color gradient that ranges from green to red, representing the likelihood of areas being classified as part of the periphery, intermediate areas, or the main core.}
    \label{fig 4}
\end{figure*}

\subsection{\label{sec:ineq}L and Population: a Proxy for Inequalities}

We compared the population living in the hexagons with the corresponding $L$: we studied the cumulative distribution of the population for different $L$. The result for all the categories and some cities is shown in Fig.~\ref{fig 3}; see the plots for all cities in Supplemental Material \cite{SM}. We noticed that each curve approaches a certain value of $L$. Generally, we observed that categories related to essential services tend to approach this value at higher $L$ compared to other categories.

As already seen, we can identify in $L=0$ a sort of benchmark: the fraction of the population living in hexagons with $L>0$ is the one that has good accessibility to POIs, being surrounded by a number of POIs per capita larger than the city average. In general, we can argue that the categories that show a more significant fraction of the population living in the hexagons with positive $L$ are the ones linked to essential services ("Transport", "Fuel and parking", "Health", etc.), while the lower fraction of population is for categories concerning consumer services and experiences ("Tourism", "Accommodation", "Catering", etc.). Looking at the maps in Fig.~\ref{fig 2}, it can be broadly inferred that the portion of the population living in hexagons with negative $L$ is mainly situated outside the city center.

\subsection{\label{sec:cores}Identifying urban areas}

\begin{figure*}
    \centering
\includegraphics[width=\textwidth]{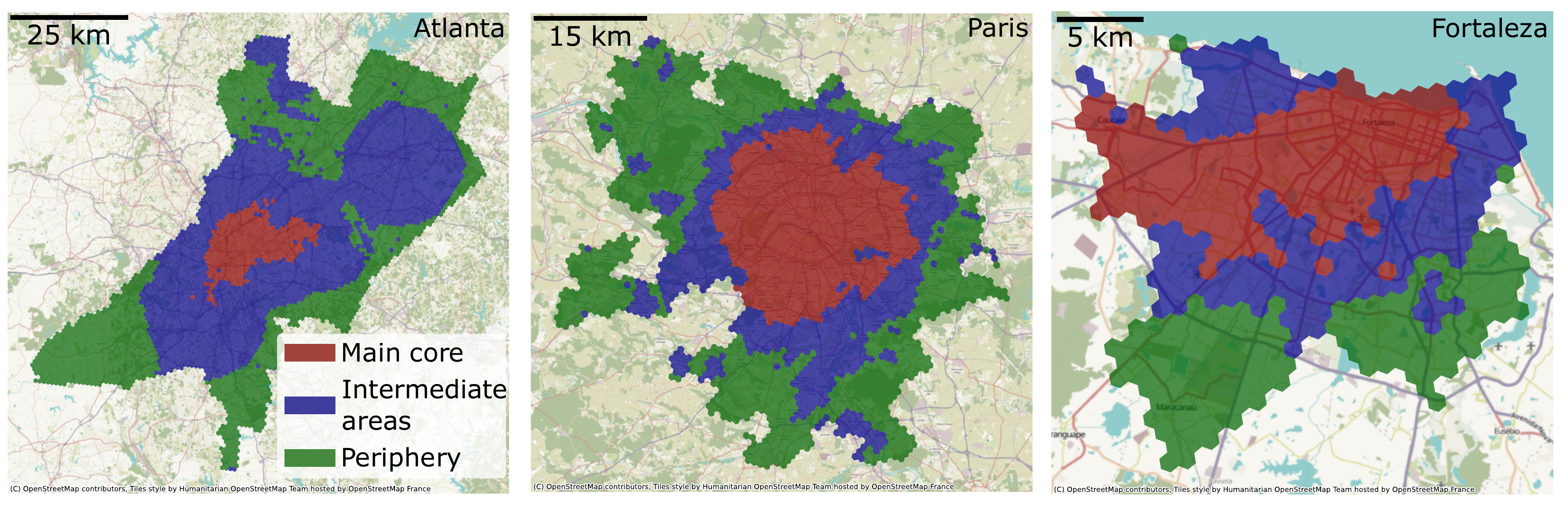}
    \caption{\textbf{Maps of core-periphery areas in Atlanta, Paris, and Fortaleza.} The maps are colored according to the outcome of the method adopted to identify the components of the urban structure. The category shown is \textit{Transport}. A large main core is observable in all cities, while the presence and extent of intermediate areas vary across cities.}
    \label{fig 5}
\end{figure*}
In our study, we analyzed the distribution of the values of $L$ for each category and city. Examples of these distributions can be seen in Fig.~\ref{fig 4}, and all plots can be found in the Supplemental Material~\cite{SM}. One common pattern we observed in these distributions is the presence of two prominent peaks: one corresponding to negative values of $L$ and another to positive values. The position of these peaks varies depending on the categories and cities.

\begin{table*}[ht]

\centering
\begin{ruledtabular}
\begin{tabular}{ccccccccccc}
	&	Atlanta	&	Barcelona	&	Bergamo	&	Bogota	&	Fortaleza	&	Milwaukee	&	Naples	&	Paris	&	Rome	&	Vienna	\\ \hline
Public	&	26.15\%	&	54.91\%	&	82.68\%	&	87.74\%	&	49.65\%	&	54.72\%	&	53.07\%	&	53.95\%	&	64.85\%	&	56.55\%	\\ 
Health	&	18.96\%	&	60.71\%	&	76.78\%	&	99.84\%	&	37.20\%	&	31.81\%	&	50.60\%	&	50.44\%	&	66.49\%	&	44.93\%	\\
Leisure	&	34.81\%	&	65.77\%	&	87.21\%	&	86.74\%	&	20.58\%	&	49.56\%	&	49.70\%	&	58.06\%	&	72.05\%	&	51.94\%	\\
Catering	&	15.44\%	&	57.98\%	&	75.45\%	&	99.84\%	&	32.09\%	&	27.05\%	&	41.62\%	&	42.36\%	&	58.22\%	&	39.90\%	\\ 
Accommodation	&	13.66\%	&	47.05\%	&	76.46\%	&	99.84\%	&	32.95\%	&	45.04\%	&	37.29\%	&	67.24\%	&	56.95\%	&	60.50\%	\\
Shopping	&	25.11\%	&	56.66\%	&	76.49\%	&	99.84\%	&	31.66\%	&	32.96\%	&	41.14\%	&	44.12\%	&	62.19\%	&	44.95\%	\\
Money	&	13.49\%	&	60.62\%	&	75.40\%	&	99.84\%	&	27.82\%	&	20.51\%	&	41.14\%	&	45.88\%	&	62.61\%	&	44.07\%	\\
Tourism	&	27.40\%	&	50.68\%	&	53.18\%	&	99.83\%	&	21.09\%	&	25.26\%	&	38.20\%	&	72.88\%	&	42.38\%	&	38.54\%	\\
Fuel and parking	&	68.03\%	&	56.02\%	&	47.53\%	&	99.84\%	&	25.41\%	&	35.19\%	&	62.32\%	&	47.06\%	&	73.70\%	&	45.81\%	\\
Transport	&	14.51\%	&	66.95\%	&	82.40\%	&	99.84\%	&	44.49\%	&	66.21\%	&	62.12\%	&	60.98\%	&	76.58\%	&	61.94\%	\\
																		
\end{tabular}
\end{ruledtabular}
\caption{Percentages of the population living in hexagons classified as core.}
\label{table core}
% \end{table*}

% \begin{table*}[ht]
\centering
\begin{ruledtabular}
\begin{tabular}{ccccccccccc}
	&	Atlanta	&	Barcelona	&	Bergamo	&	Bogota	&	Fortaleza	&	Milwaukee	&	Naples	&	Paris	&	Rome	&	Vienna	\\ \hline
Public	&	51.95\%	&	25.22\%	&	-	&	12.07\%	&	-	&	17.10\%	&	-	&	20.68\%	&	-	&	-	\\ 
Health	&	73.49\%	&	27.71\%	&	-	&	-	&	35.04\%	&	40.58\%	&	-	&	19.73\%	&	24.53\%	&	-	\\
Leisure	&	38.91\%	&	22.67\%	&	-	&	13.07\%	&	51.93\%	&	40.31\%	&	-	&	31.56\%	&	23.14\%	&	25.71\%	\\
Catering	&	73.38\%	&	28.85\%	&	-	&	-	&	39.61\%	&	37.71\%	&	-	&	38.78\%	&	-	&	-	\\ 
Accommodation	&	75.29\%	&	38.53\%	&	-	&	-	&	-	&	31.07\%	&	-	&	-	&	34.14\%	&	-	\\
Shopping	&	71.39\%	&	29.44\%	&	-	&	-	&	43.24\%	&	42.37\%	&	-	&	28.59\%	&	28.99\%	&	-	\\
Money	&	79.43\%	&	26.11\%	&	-	&	-	&	40.24\%	&	63.40\%	&	-	&	24.13\%	&	28.10\%	&	-	\\
Tourism	&	64.91\%	&	32.21\%	&	-	&	-	&	31.19\%	&	56.02\%	&	-	&	-	&	45.98\%	&	-	\\
Fuel and parking	&	28.21\%	&	28.14\%	&	34.99\%	&	-	&	32.26\%	&	30.65\%	&	-	&	29.36\%	&	18.74\%	&	-	\\
Transport	&	59.05\%	&	24.04\%	&	-	&	-	&	36.57\%	&	18.78\%	&	-	&	22.97\%	&	18.81\%	&	-	\\
																			
\end{tabular}
\end{ruledtabular}
 \caption{Percentages of the population living in hexagons classified as intermediate areas.}

\label{table intermediate}

\centering
\begin{ruledtabular}
\begin{tabular}{ccccccccccc}
	&	Atlanta	&	Barcelona	&	Bergamo	&	Bogota	&	Fortaleza	&	Milwaukee	&	Naples	&	Paris	&	Rome	&	Vienna	\\ \hline
Public	&	21.90\%	&	19.88\%	&	17.32\%	&	0.19\%	&	50.35\%	&	28.18\%	&	46.93\%	&	25.36\%	&	35.15\%	&	43.45\%	\\ 
Health	&	7.55\%	&	11.58\%	&	23.22\%	&	0.16\%	&	27.76\%	&	27.60\%	&	49.40\%	&	29.84\%	&	8.98\%	&	55.07\%	\\
Leisure	&	26.28\%	&	11.56\%	&	12.79\%	&	0.19\%	&	27.49\%	&	10.14\%	&	50.30\%	&	10.38\%	&	4.81\%	&	22.34\%	\\
Catering	&	11.18\%	&	13.17\%	&	24.55\%	&	0.16\%	&	28.30\%	&	35.24\%	&	58.38\%	&	18.86\%	&	41.78\%	&	60.10\%	\\ 
Accommodation	&	11.05\%	&	14.42\%	&	23.54\%	&	0.16\%	&	67.05\%	&	23.89\%	&	62.71\%	&	32.76\%	&	8.90\%	&	39.50\%	\\
Shopping	&	3.49\%	&	13.90\%	&	23.51\%	&	0.16\%	&	25.10\%	&	24.68\%	&	58.86\%	&	27.30\%	&	8.82\%	&	55.05\%	\\
Money	&	7.08\%	&	13.27\%	&	24.60\%	&	0.16\%	&	31.94\%	&	16.09\%	&	58.86\%	&	30.00\%	&	9.28\%	&	55.93\%	\\
Tourism	&	7.70\%	&	17.11\%	&	46.82\%	&	0.17\%	&	47.72\%	&	18.72\%	&	61.80\%	&	27.12\%	&	11.65\%	&	61.46\%	\\
Fuel and parking	&	3.75\%	&	15.84\%	&	17.48\%	&	0.16\%	&	42.34\%	&	34.16\%	&	37.68\%	&	23.58\%	&	7.56\%	&	54.19\%	\\
Transport	&	26.44\%	&	9.01\%	&	17.60\%	&	0.16\%	&	18.94\%	&	15.01\%	&	37.88\%	&	16.06\%	&	4.61\%	&	38.06\%	\\
																		
\end{tabular}
\end{ruledtabular}
\caption{Percentages of the population living in hexagons classified as periphery.}
\label{table periphery}
\end{table*}

The observed peaks are connected to different properties of accessibility in the city structure. 
To identify qualitative differences among different areas, we used a Gaussian Mixture Model for clustering hexagons in separate peaks of the $L$ distribution, adopting the Bayesian Information Criterion to determine the optimal number of Gaussian components. We then assigned each hexagon on the map to a specific Gaussian component based on the highest probability of belonging to that component. In our analysis, we found two or three distinct components. We classified the cluster with the highest mean of the Gaussian component as the main core, while hexagons associated with the Gaussian component having the lowest mean were classified as periphery. If a third component was present, hexagons related to it were categorized as intermediate areas, which can be surrounding centers or detached smaller centers.  In Fig.~\ref{fig 4}, we provide examples of the outcomes of our method for Rome for different categories of services, and in Fig.~\ref{fig 5}, we show the outcome for other cities. See all maps in the Supplemental Material~\cite{SM}. In summary, we can categorize the areas as follows:

\begin{itemize}

    \item A \textit{primary core} is always found as the spatial cluster with positive highest values of L and with the largest mean of the corresponding Gaussian component. There is a consistent presence of a large main core in all cities and categories of services. This main core acts as the functional hub of the city and is distinguished by a large area of significantly good accessibility, as the depletion zone emerges at greater distances than other areas of the city. In most cases, the main core aligns with the city geographical center, while the extent of the core area depends on the category. Maps of categories related to essential services, such as \textit{Transport} or \textit{Health}, depict a larger area compared to other categories like \textit{Accommodation} and \textit{Tourism}. 

    \item The \textit{periphery} includes all cells with negative values of $L$ belonging to the Gaussian component with the lowest mean. It extends further for categories related to consumer services and experiences, such as \textit{Tourism}, \textit{Accommodation}, and \textit{Catering}, due to their high spatial concentration. 
    
    \item The \textit{intermediate area} is a spatial cluster characterized by $L$ values with the highest probability of belonging to a third Gaussian component whose mean lies between the peripheral and the main core regions. The intermediate areas encompass zones around the city center or disconnected from it but are characterized by a consistent number of services or a concentration of POIs of a specific category. For instance, in the case of Rome (Fig.~\ref{fig 4}), the archaeological sites of \textit{Ostia Antica} and \textit{Parco di Veio} justify two intermediate areas for the \textit{Tourism} category.
    
\end{itemize}

To gain a better understanding of the differences between the categories, we can refer to Fig.~\ref{fig 4}, which compares the \textit{Catering} and \textit{Transport} categories. This comparison shows that the intermediate areas in \textit{Transport} disappear and are classified as periphery in the \textit{Catering} category. All hexagons belonging to these clusters are characterized by a negative or small positive value of "L", indicating a depletion of services and low accessibility to services. 

To analyze the relationship with the population distribution, we calculated the percentage of people residing in hexagons classified as core, intermediate areas, and periphery. The results are shown in Tables \ref{table core}, \ref{table intermediate}, and \ref{table periphery}. Generally, a large part of the population is concentrated in the core, although a considerable part of the population is spread between the intermediate areas (when present) and the periphery. An exception is Atlanta, where most of the population is in hexagons classified as intermediate areas, indicating a different spatial distribution of services and population. We also weighted the histogram of $L$ values with the population in each hexagon. We observed that the peaks whose height becomes smaller are related to the peripheral area: the hexagons characterized by negative values of $L$ show a low number of people living there. The peaks associated with the functional core and the intermediate areas are the most populated parts of the city. The results and a more detailed discussion can be found in the Supplemental Material~\cite{SM}.

\section{\label{sec:discussion}Discussion}

Identifying the central cores of activity in a city and its surrounding areas is crucial for understanding urban development. Cities are often treated as a single entity, overlooking empty spaces and spatial inequalities. Pinpointing key activity centers provides policymakers with more accurate data on land use, human behavior, and urban growth, leading to more effective policy recommendations.

This study introduces a novel approach to understanding city structure by employing a new accessibility metric that encompasses accessibility to close and far services. We applied the radial distribution function to quantitatively analyze urban layouts by assessing the likelihood of encountering Points of Interest (POIs) at various distances, relative to a uniform distribution. Our findings reveal distinct accessibility patterns within urban landscapes. Essential services, such as healthcare and public facilities, exhibited broader accessibility ranges (higher $L$) compared to consumer services, indicating a more even distribution across the urban landscape. This suggests that urban policies prioritize equitable access to essential services, ensuring that residents, regardless of location, have better access to these critical resources. In contrast, consumer services are more concentrated in specific commercial zones, reflecting economic strategies aimed at maximizing efficiency and catering to consumer demand patterns.

Building on our definition of accessibility, we developed a technique to identify a city's main core, intermediate areas, and periphery. Cities like Vienna and Naples exhibit a clear core-periphery structure, likely shaped by historical patterns where essential services are concentrated in the central urban area, while outlying areas remain more residential with fewer services. In contrast, cities like Paris and Rome display not only a central core but also well-defined intermediate zones, likely due to their larger size and more complex urban histories. These intermediate areas emerged to serve growing populations, ensuring broader access to both essential and consumer services across different parts of the city.

Our analysis also incorporated population data, revealing which areas with low accessibility have also a high population density. Using our accessibility metric, we can quantify the percentage of individuals living in low-accessibility areas, offering a preliminary indicator of spatial inequalities. These findings invite further exploration through detailed spatial analyses to uncover the socio-economic factors driving disparities in service access and population distribution across urban areas.

Looking ahead to further development of this work, incorporating mobility data would offer a richer perspective. This integration could enable comparative studies between the static spatial distribution of Points of Interest and the dynamic patterns of human mobility, providing deeper insights into urban structure and accessibility.

In conclusion, we have developed a method to analyze urban meso-structures using a novel accessibility metric. Combined with population data, this approach reveals key insights into urban inequalities and accessibility patterns. These findings can equip policymakers with tools to address disparities and enhance urban planning efforts.

\appendix*

\section{\label{sec:data}Material}
We used the OECD's functional urban area definition \cite{dijkstra2019eu} for all cities, except for Fortaleza, for which we took the GHS-FUA definition \cite{florczyk2019ghs}. For every city we restricted our analysis to the largest connected component, deleting disconnected areas if present.  

We used a hexagonal grid that evenly divides the city to study and compare individual grid cells. The segmentation of urban areas into hexagons is accomplished through the use of the Python package \texttt{h3} \cite{h3_package}. The resolution of the hexagons utilized is such that the average area of each hexagon is about 0.7 km$^2$. 

The data about Points of Interest was extracted from OpenStreetMap \cite{osm}. A comprehensive evaluation of the dataset was conducted to assess its utility and reliability. From the characterization provided by a system of codes of OpenStreetMap, ten categories of POIs were identified:

\begin{enumerate}[label=(\arabic*)] 
    \item \emph{Public}: police station, fire station, post office, library, town hall, university, school, public telephone booth, etc.
    \item \emph{Health}: pharmacy, hospital, medical centre, medical practice, dentist's practice, veterinary, etc.
    \item \emph{Leisure}: theatre, nightclub, cinema, park, playground, swimming pool, stadium, etc.
    \item \emph{Catering}: restaurant, fast food, cafe, pub, bar, food court, etc.
    \item \emph{Accommodation}: hotel, motel, bed and breakfast, guesthouse, hostel, campsite, etc.
    \item \emph{Shopping}: supermarket, bakery, mall, department store, clothes store, florist, bookshop, toy store, car garage, car sharing station, laundry, etc.
    \item \emph{Money}: bank, atm
    \item \emph{Tourism}: tourist attraction, tourist information, museum, monument, memorial, castle, ruins, picnic site, zoo, viewpoint, etc.
    \item \emph{Fuel and parking}: gas station, service area, car park, bicycle park, etc.
    \item \emph{Transport}: bus stop, tram stop, railway station, bus station, taxi rank, airport, etc.
\end{enumerate}

To compare with the spatial distribution of the resident population we used data from WorldPop \cite{worldpop} that contains raster images composed of a grid of pixels, each of which has a numerical value that corresponds to the population in that cell. We employed the 100m population grid adjusted to municipal UN population estimates ~\cite{bondarenko2020census}.

\providecommand{\noopsort}[1]{}\providecommand{\singleletter}[1]{#1}%

\end{document}

\maketitle

\tableofcontents

\newpage

\section{$\sigma_{g(r)}$ versus $\Delta r$}

\begin{figure}[H]
    \centering
    \includegraphics[width=\textwidth]{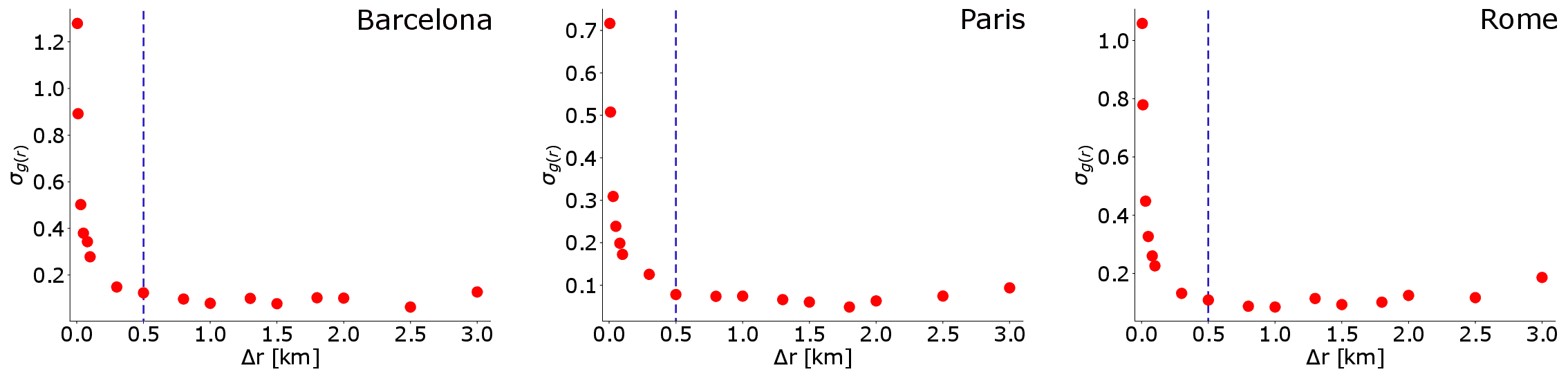}
    \caption{$\sigma_{g(r)}$ vs $\Delta r$ for the category "Transport". The plots are showed for Barcelona, Paris and Rome. The blue dashed line corresponds to the chosen $\Delta r$.}
    \label{stdev}
\end{figure}

Before computing the radial distribution function $g(r)$, we conducted an analysis to determine the optimal value of $\Delta r$. If Points of Interest are uniformly distributed within a city, $g(r)$ is equal one for all selected $r$. We redistributed POIs randomly within urban areas according to a uniform distribution and we observed fluctuations in the calculated $g(r)$. These fluctuations varied with $\Delta r$ and we chose the $\Delta r$ that minimized these fluctuations. We calculated $g(r)$ for randomly distributed POIs using different $\Delta r$ values ranging from 0.005 km to 3 km. For each $\Delta r$ value, we computed the standard deviation $\sigma_{g(r)}$ of the $g(r)$. The resulting $\sigma_{g(r)}$ as a function of $\Delta r$ is shown in Fig. \ref{stdev} for some cities and for the category "Transport".

\section{Accessibility values}

The figures presented in this section illustrate maps of $L$ for all cities and categories. $L$ is derived from the radial distribution function and it serves as a metric for assessing accessibility, focusing on the transition from distances with clustering to those without and vice versa. This division is designated by positive and negative values of $L$. It is clear that essential services and facilities necessary for urban living have a more spread accessibility and higher values of $L$ compared to categories regarding consumer services and experiences.

\begin{figure}[H]
    \centering
    \includegraphics[width=\textwidth]{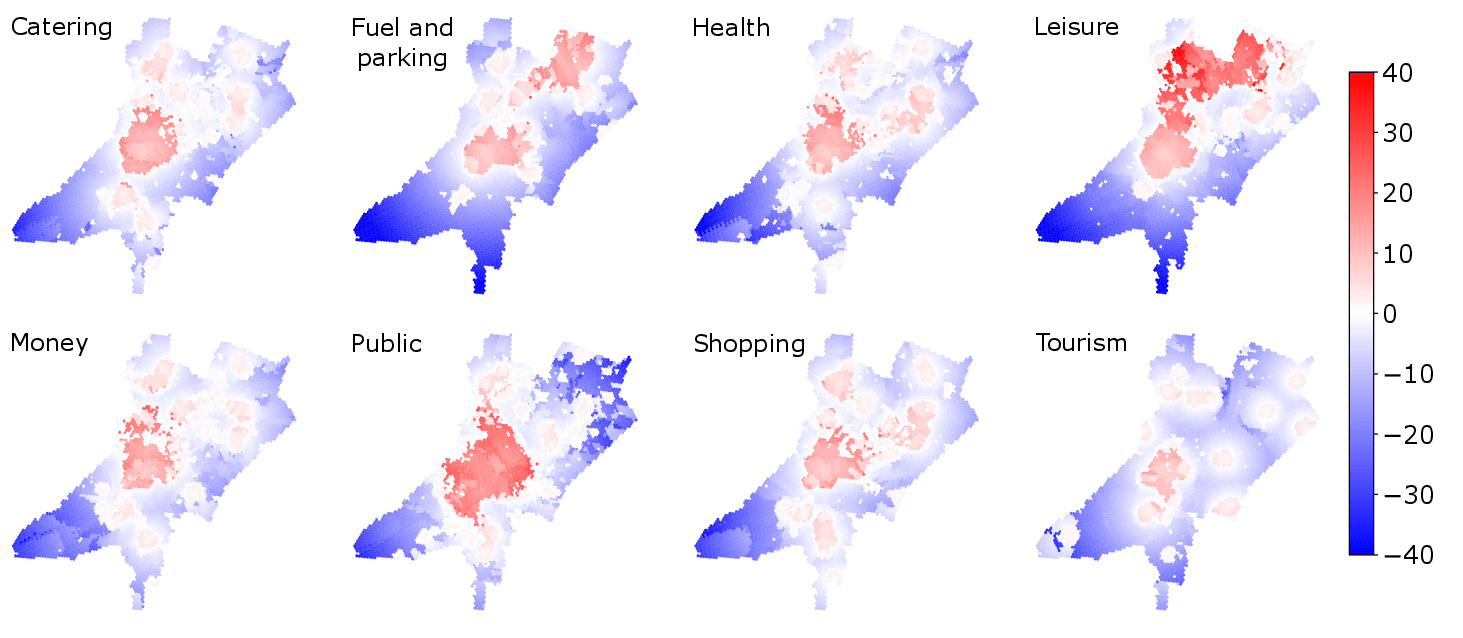}
    \caption{Maps of Atlanta representing the $L$ values for all the categories. }
\end{figure}

\begin{figure}[H]
    \centering
    \includegraphics[width=\textwidth]{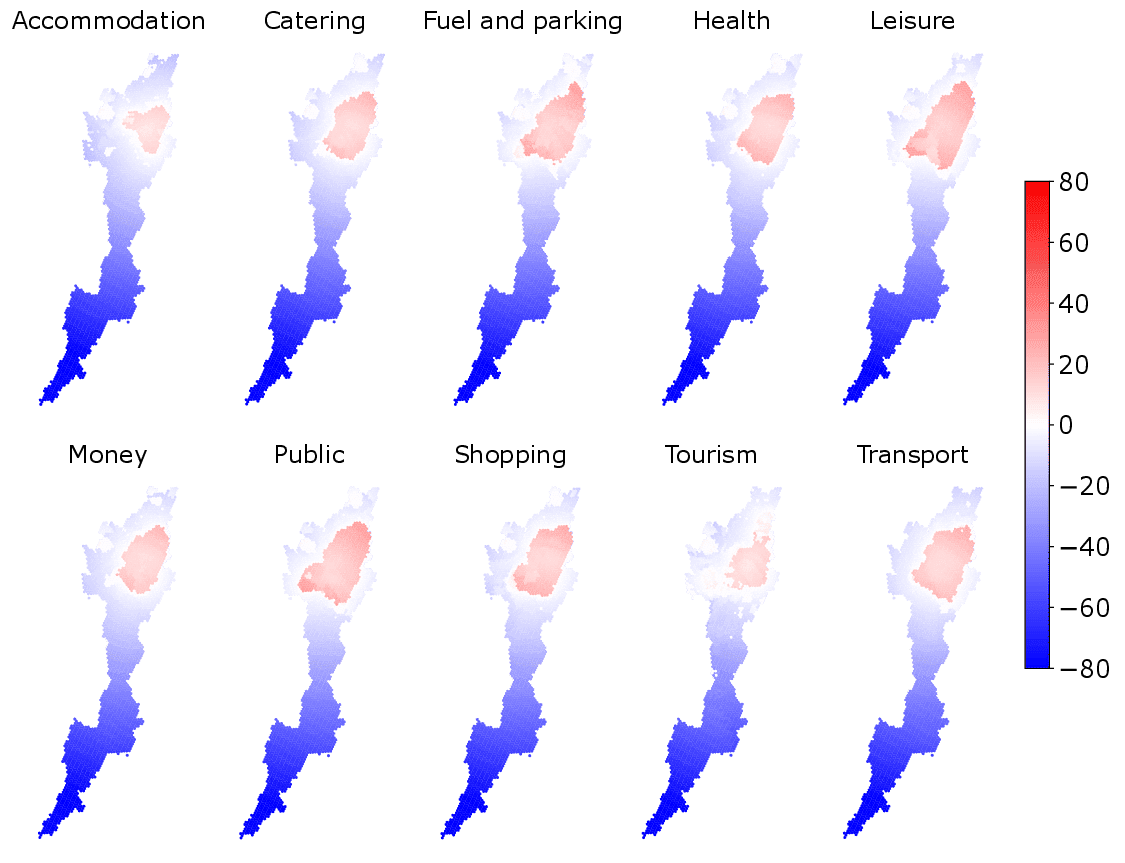}
    \caption{Maps of Bogota representing the $L$ values for all the categories. }
\end{figure}

\begin{figure}[H]
    \centering
    \includegraphics[width=\textwidth]{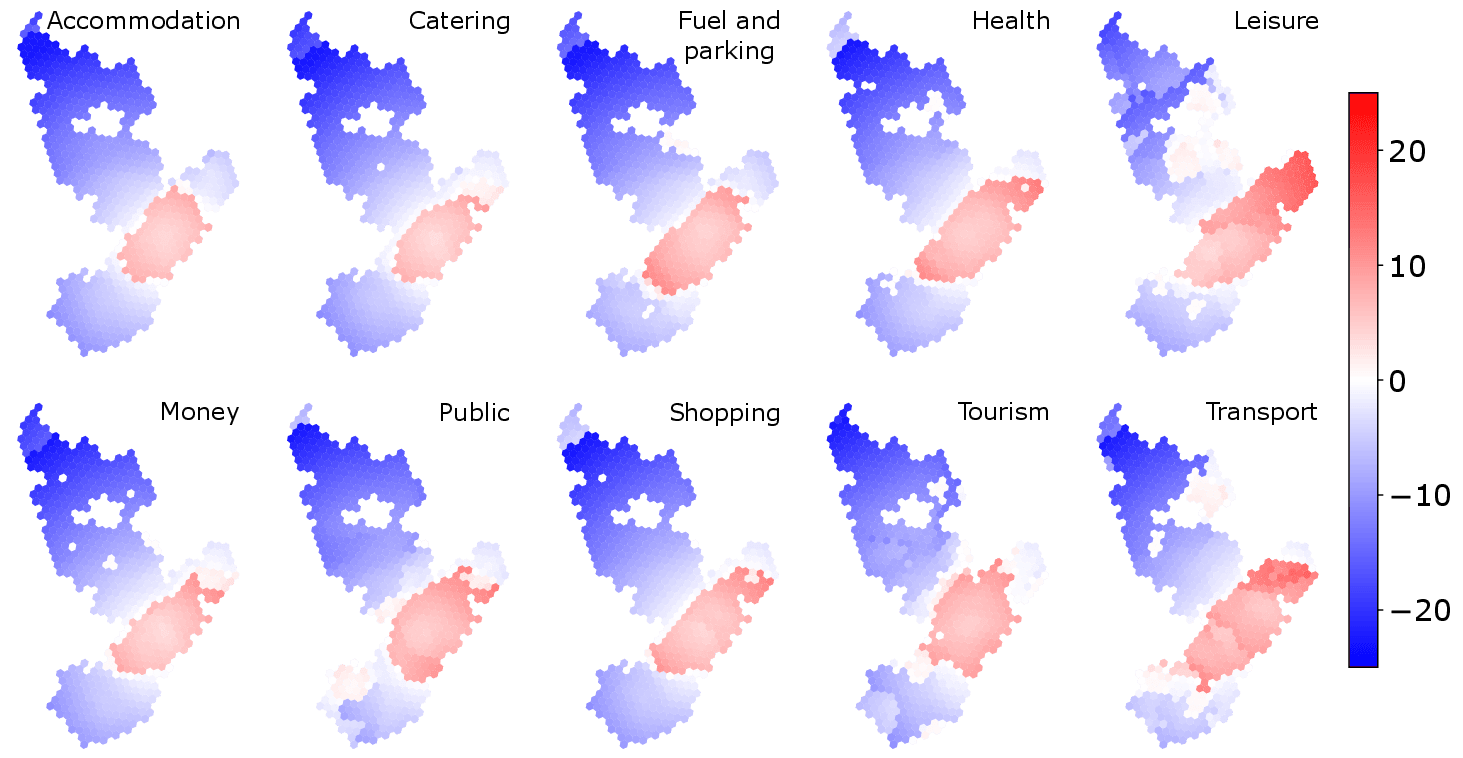}
    \caption{Maps of Barcelona representing the $L$ values for all the categories.}
\end{figure}

\begin{figure}[H]
    \centering
    \includegraphics[width=\textwidth]{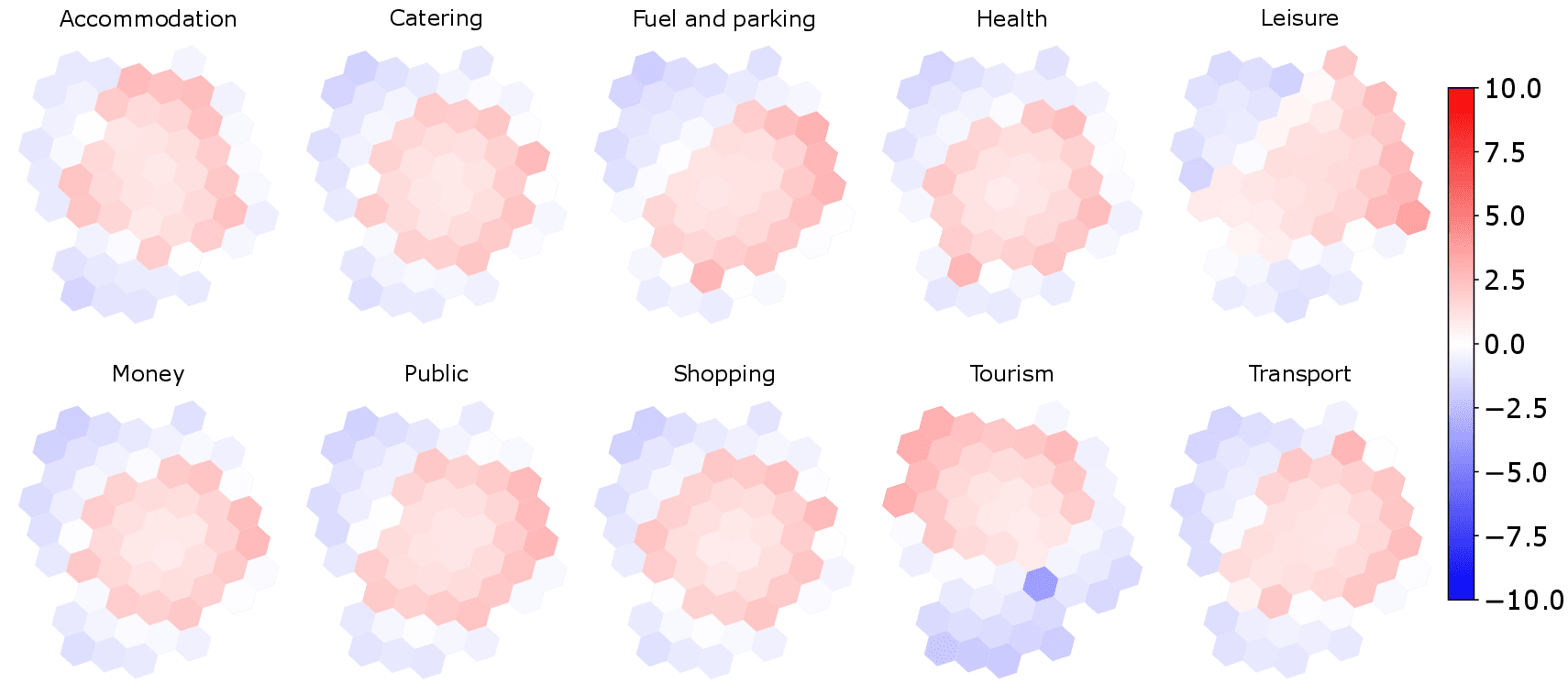}
    \caption{Maps of Bergamo representing the $L$ values for all the categories.}
\end{figure}

\begin{figure}[H]
    \centering
    \includegraphics[width=\textwidth]{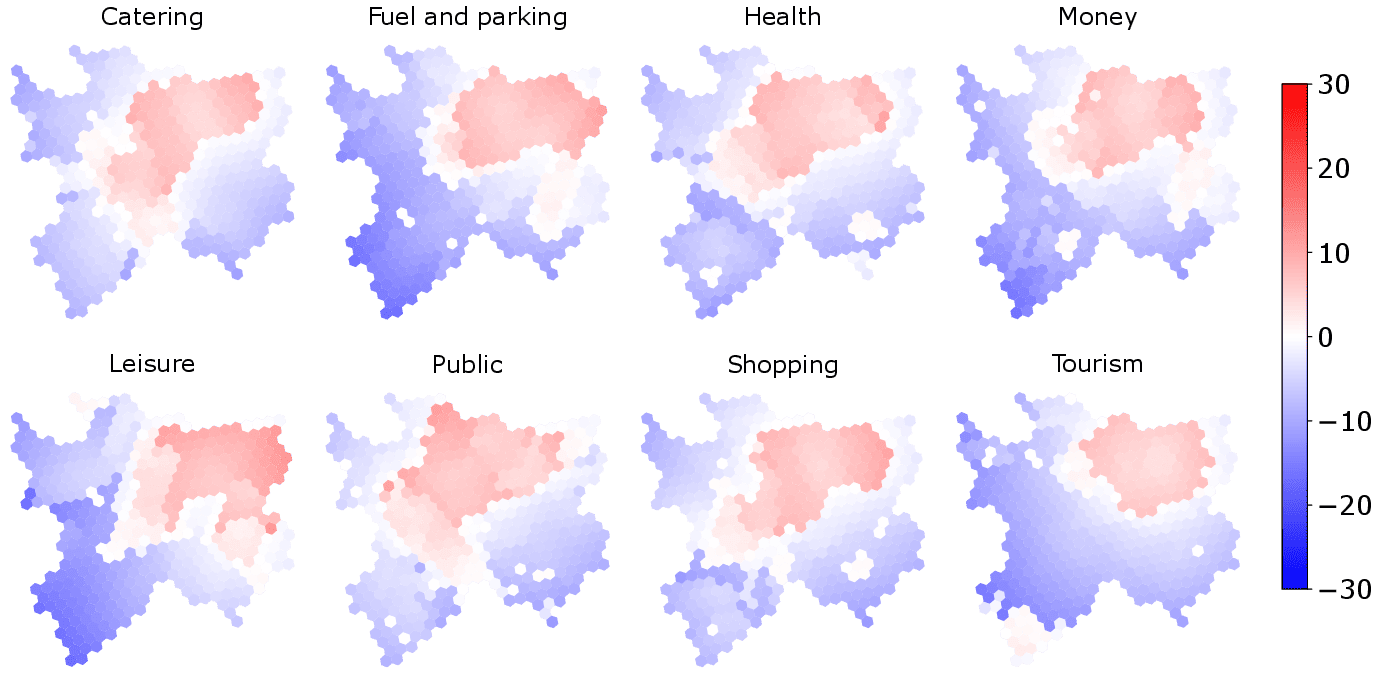}
    \caption{Maps of Fortaleza representing the $L$ values for all the categories. }
\end{figure}

\begin{figure}[H]
    \centering
    \includegraphics[width=\textwidth]{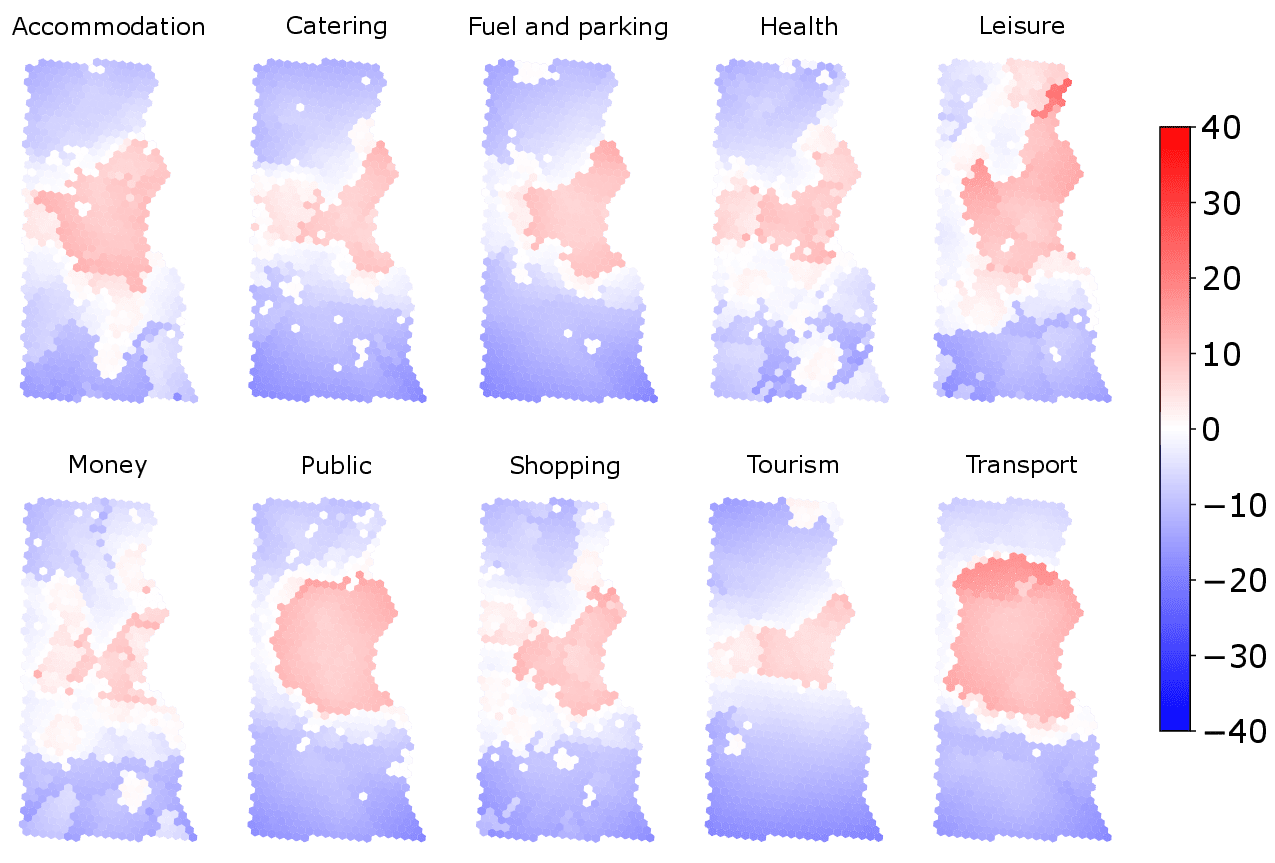}
    \caption{Maps of Milwakee representing the $L$ values for all the categories.}
\end{figure}

\begin{figure}[H]
    \centering
    \includegraphics[width=\textwidth]{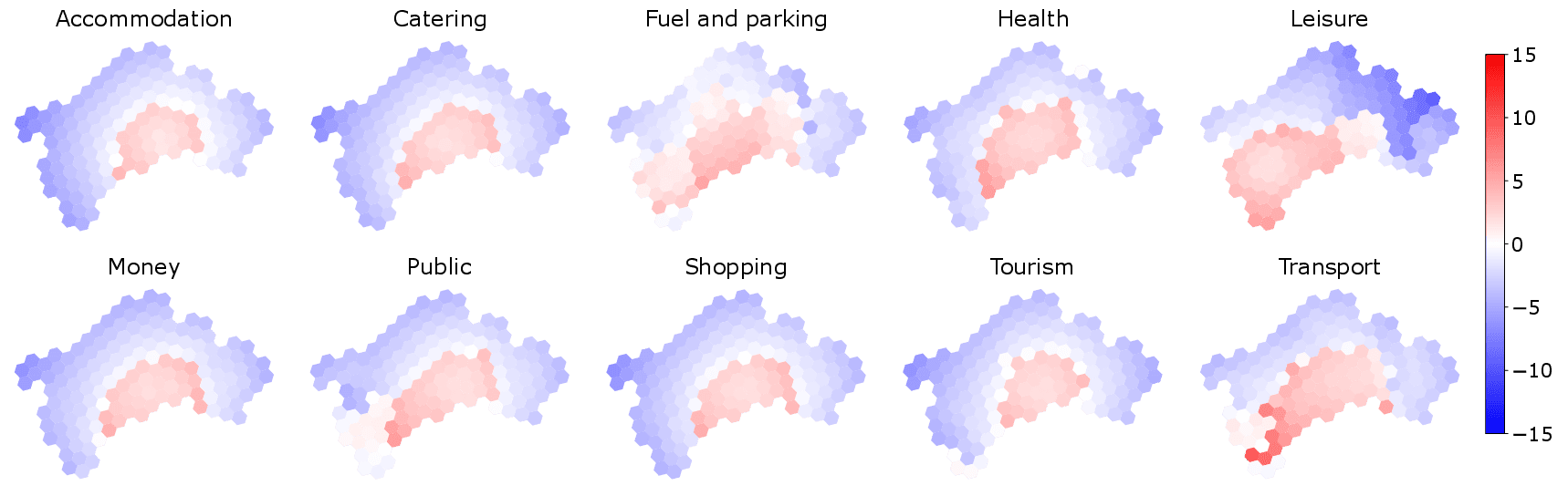}
    \caption{Maps of Naples representing the $L$ values for all the categories.}
\end{figure}

\begin{figure}[H]
    \centering
    \includegraphics[width=\textwidth]{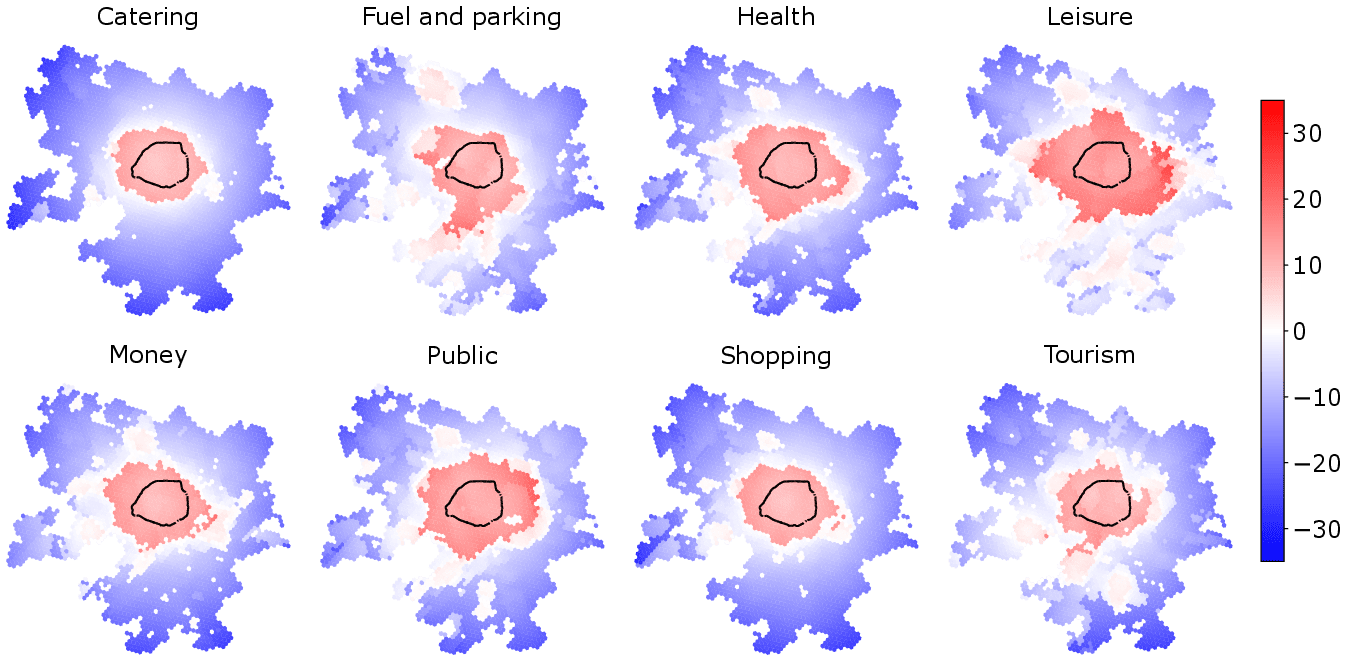}
    \caption{Maps of Paris representing the $L$ values for all the categories. The black line in the maps represents Boulevard Périphérique. }
\end{figure}

\begin{figure}[H]
    \centering
    \includegraphics[width=\textwidth]{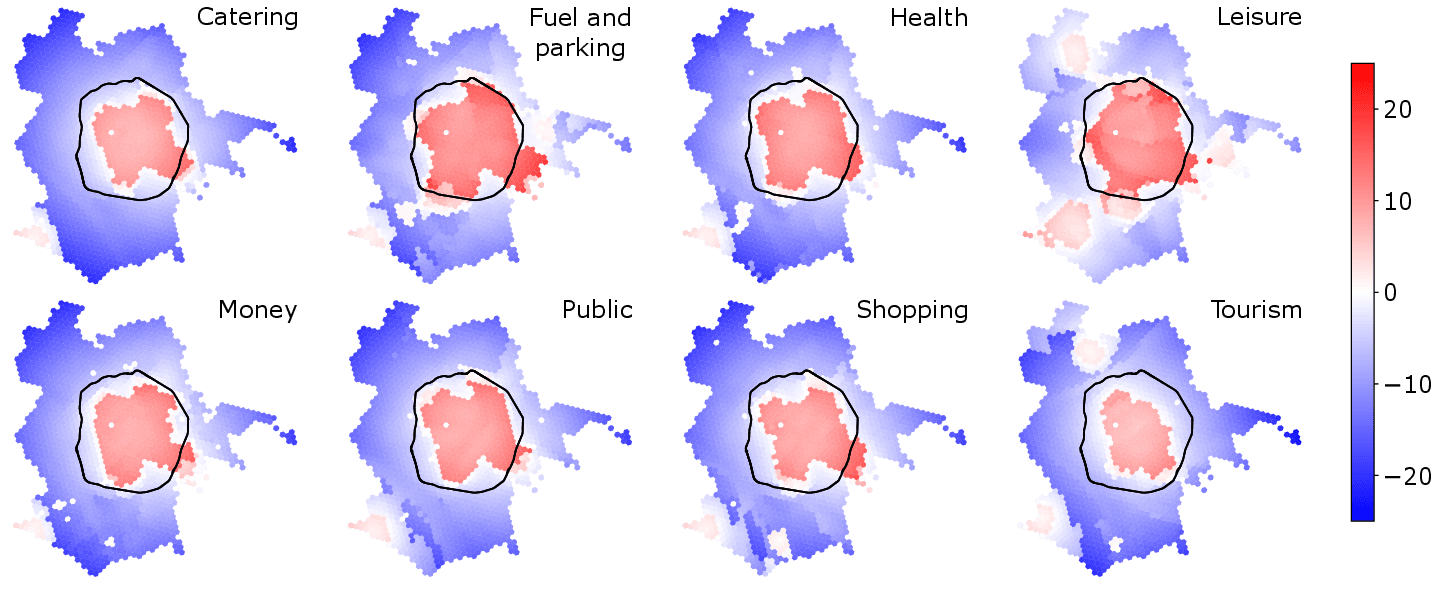}
    \caption{Maps of Rome representing the $L$ values for all the categories. The black line in the Rome maps represents Grande Raccordo Anulare.}
\end{figure}

\begin{figure}[H]
    \centering
    \includegraphics[width=\textwidth]{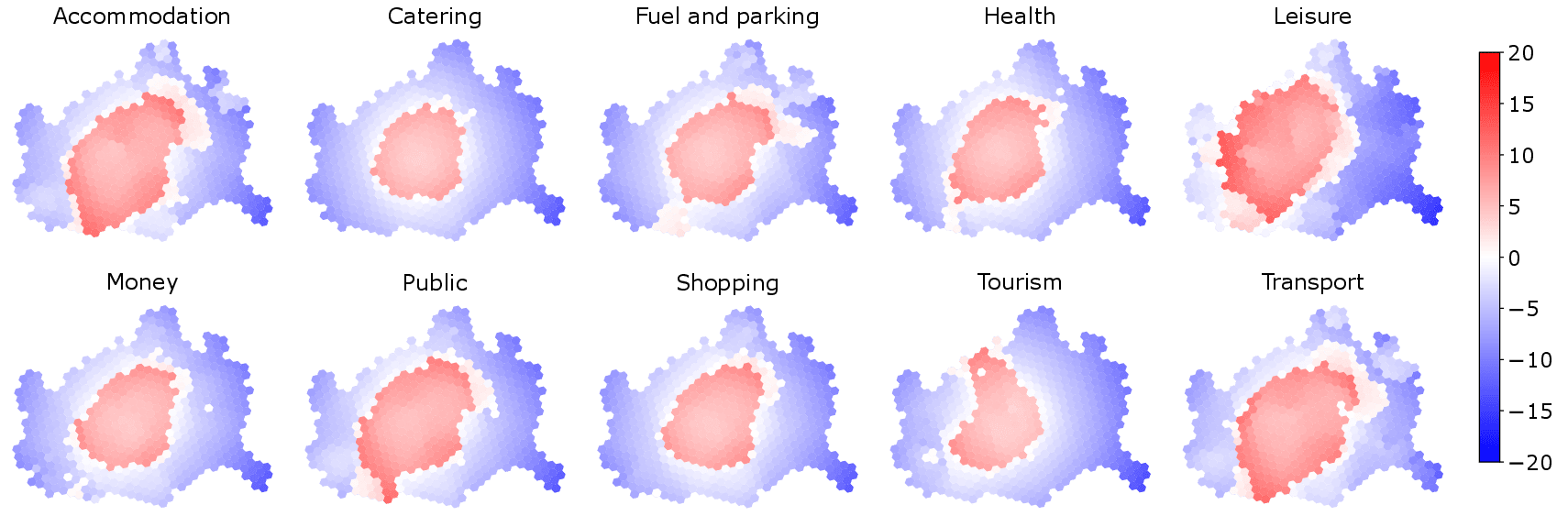}
    \caption{Maps of Vienna representing the $L$ values for all the categories. }
\end{figure}

\newpage

\section{$f_p$ vs $L$}

\begin{figure}[H]
    \centering
    \includegraphics[width=\textwidth]{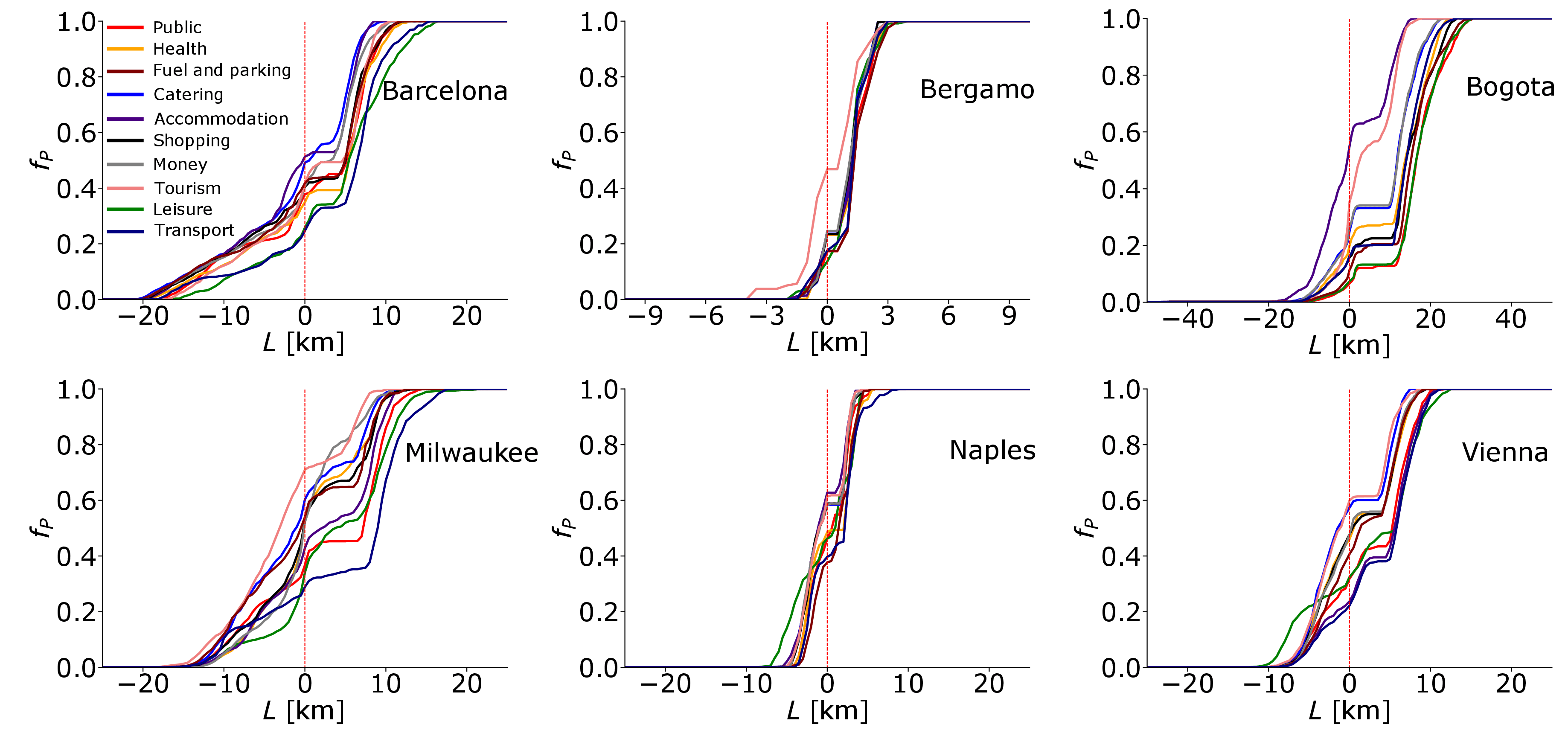}
    \caption{Cumulative distribution of population $f_P$ with the varying $L$ for Barcelona, Bergamo, Bogota, Milwaukee, Naples, Vienna.}
    \label{cumulative}
\end{figure}

Figure \ref{cumulative} illustrates the cumulative distribution of the population across different $L$. This distribution provides insights into how the population is distributed concerning varying levels of accessibility. The fraction of population at $L=0$ indicates the percentage of people living in hexagons characterized by poor accessibility. This can be seen as a measure of inequalities.

\section{Distributions of $L$ values}

In this section we show the distribution of $L$ for each city and category. We can see that depending from the city there are two or three peaks: one at negative $L$, one at positive $L$ and another near $L=0$.

\begin{figure}[H]
    \centering
\includegraphics[width=\textwidth]{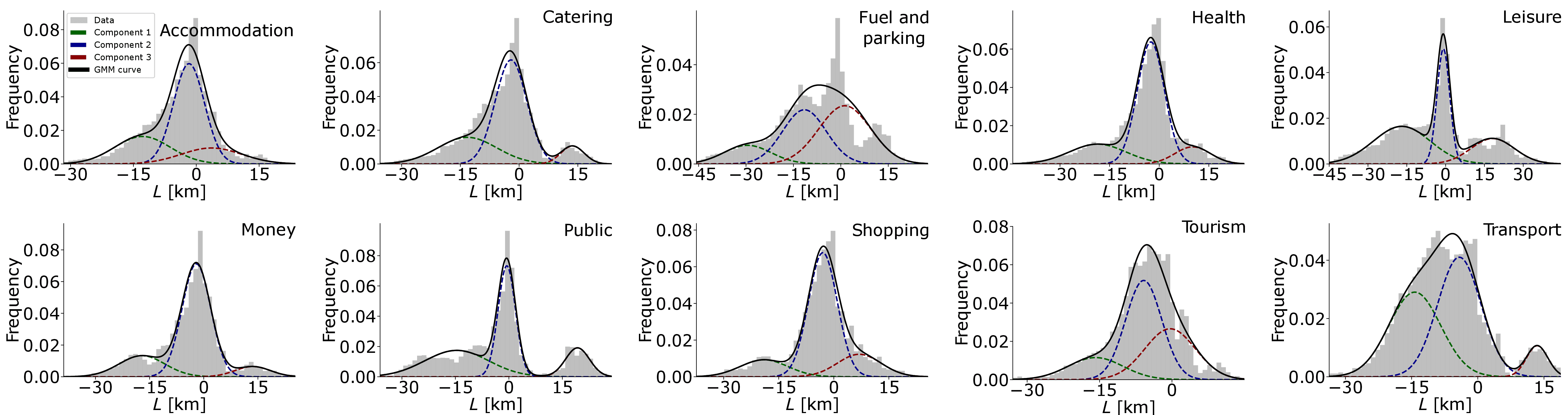}
    \caption{Histograms of $L$ values of all categories for Atlanta.}
\end{figure}

\begin{figure}[H]
    \centering
\includegraphics[width=\textwidth]{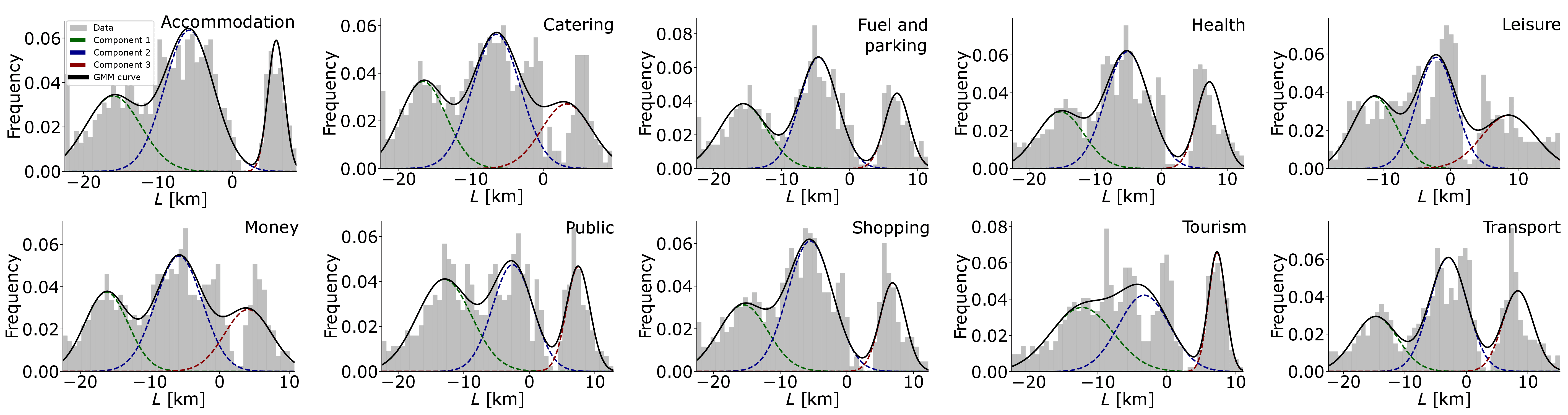}
    \caption{Histograms of $L$ values of all categories for Barcelona.}
\end{figure}

\begin{figure}[H]
    \centering
\includegraphics[width=\textwidth]{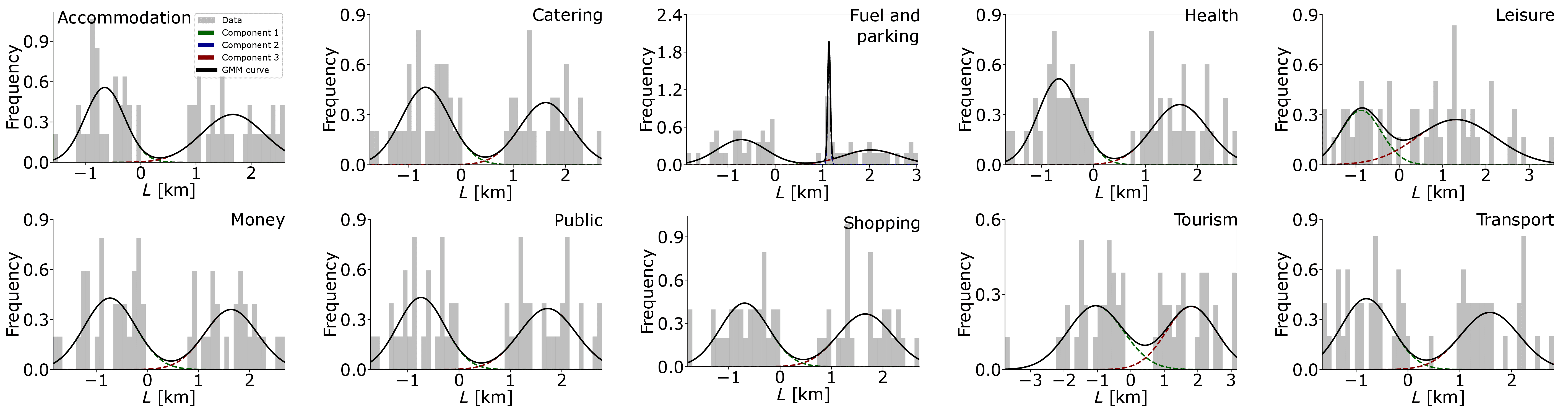}
    \caption{Histograms of $L$ values of all categories for Bergamo.}
\end{figure}

\begin{figure}[H]
    \centering
\includegraphics[width=\textwidth]{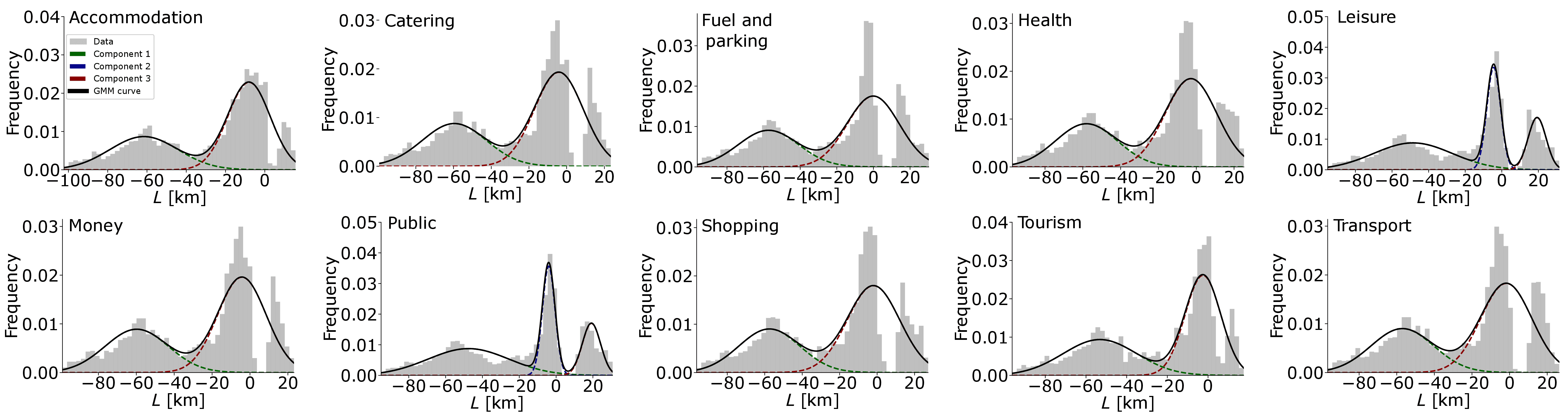}
    \caption{Histograms of $L$ values of all categories for Bogota.}
\end{figure}

\begin{figure}[H]
    \centering
\includegraphics[width=\textwidth]{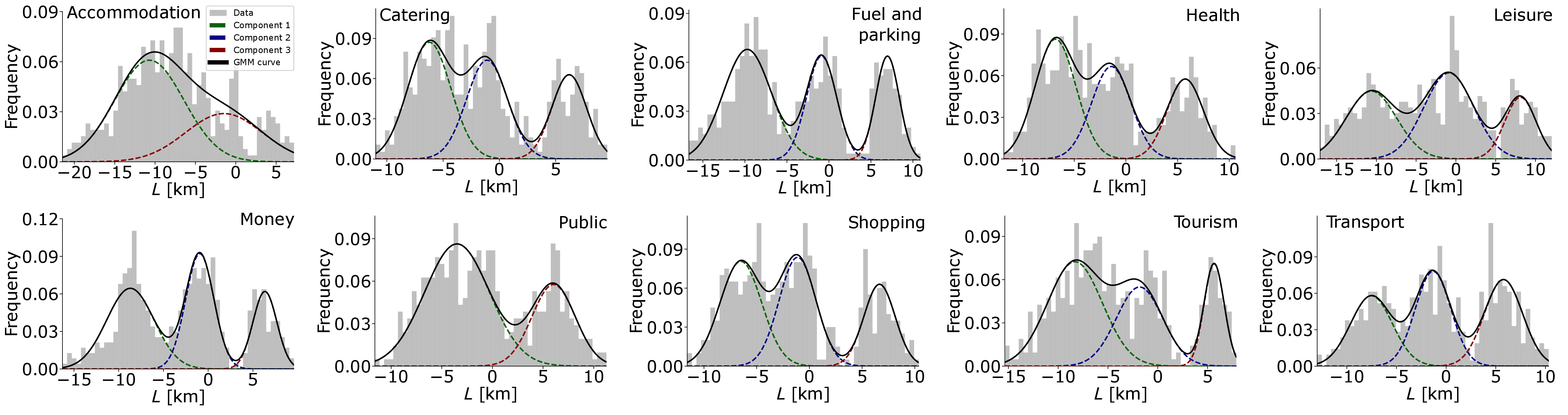}
    \caption{Histograms of $L$ values of all categories for Fortaleza.}
\end{figure}

\begin{figure}[H]
    \centering
\includegraphics[width=\textwidth]{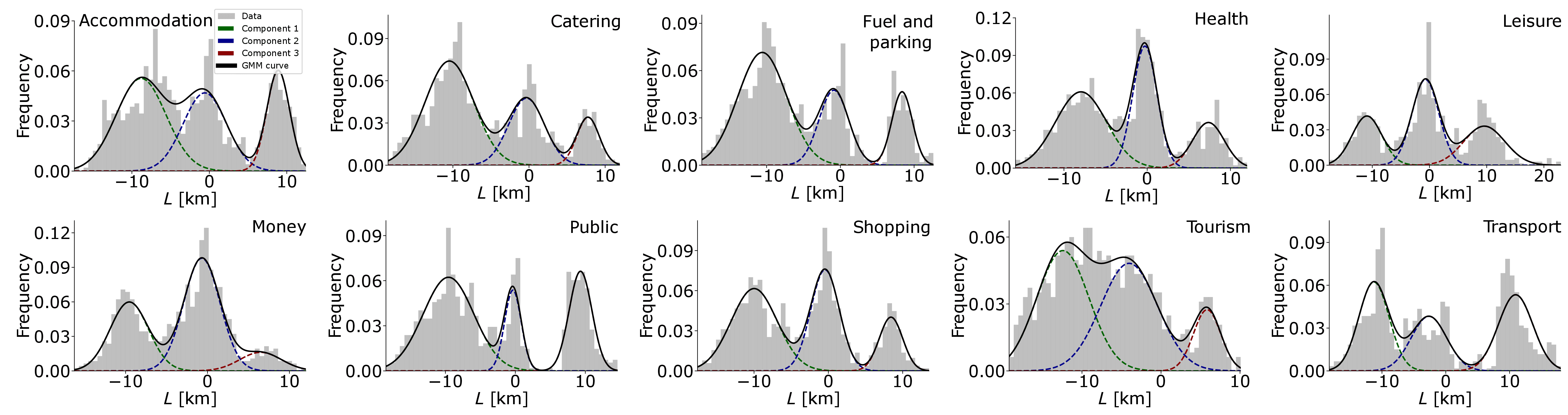}
    \caption{Histograms of $L$ values of all categories for Milwaukee.}
\end{figure}

\begin{figure}[H]
    \centering
\includegraphics[width=\textwidth]{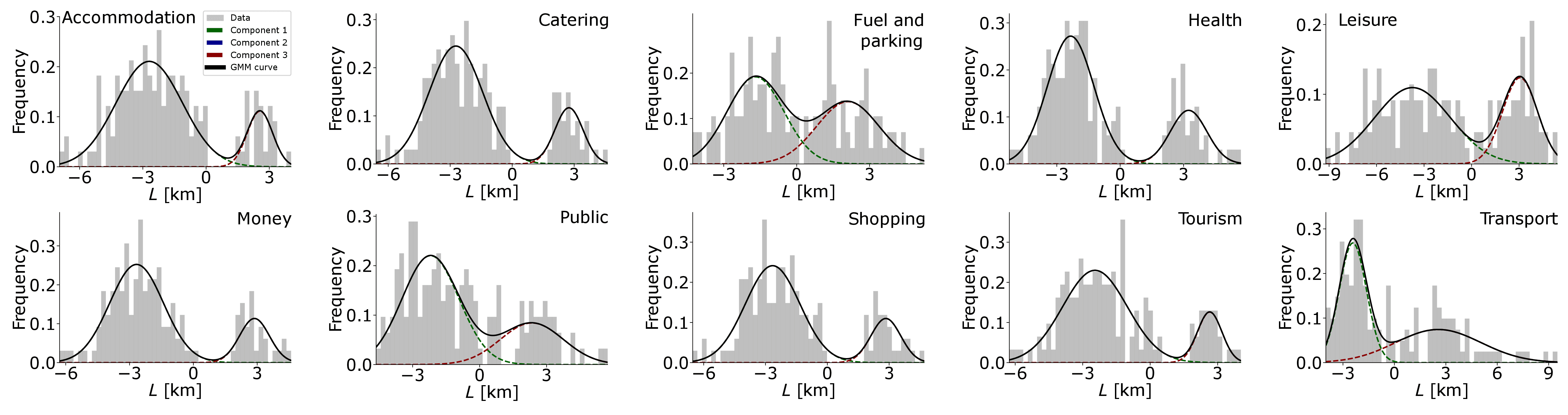}
    \caption{Histograms of $L$ values of all categories for Naples.}
\end{figure}

\begin{figure}[H]
    \centering
\includegraphics[width=\textwidth]{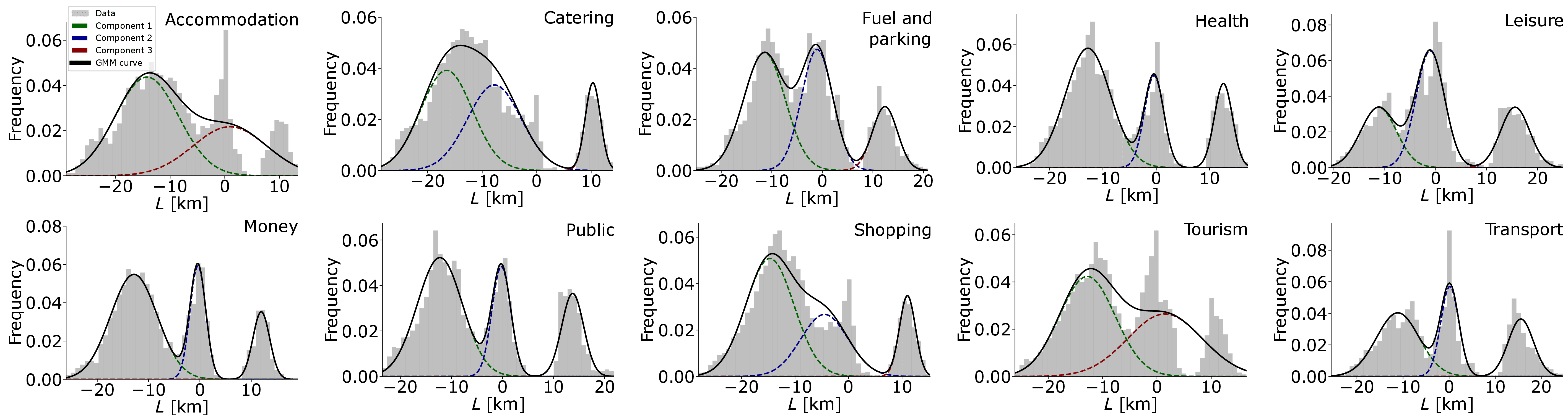}
    \caption{Histograms of $L$ values of all categories for Paris.}
\end{figure}

\begin{figure}[H]
    \centering
\includegraphics[width=\textwidth]{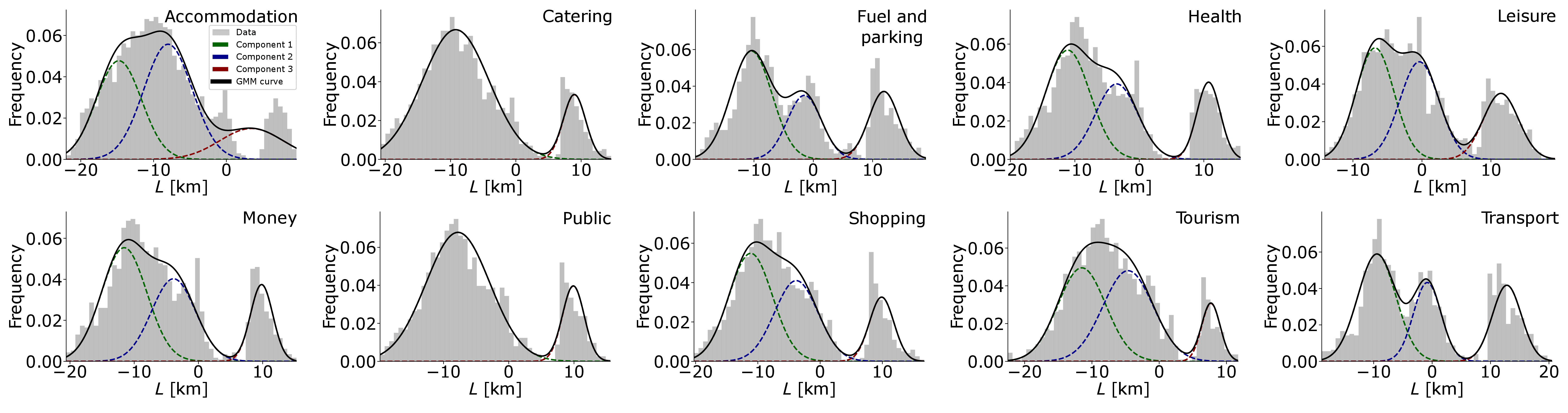}
    \caption{Histograms of $L$ values of all categories for Rome.}
\end{figure}

\begin{figure}[H]
    \centering
\includegraphics[width=\textwidth]{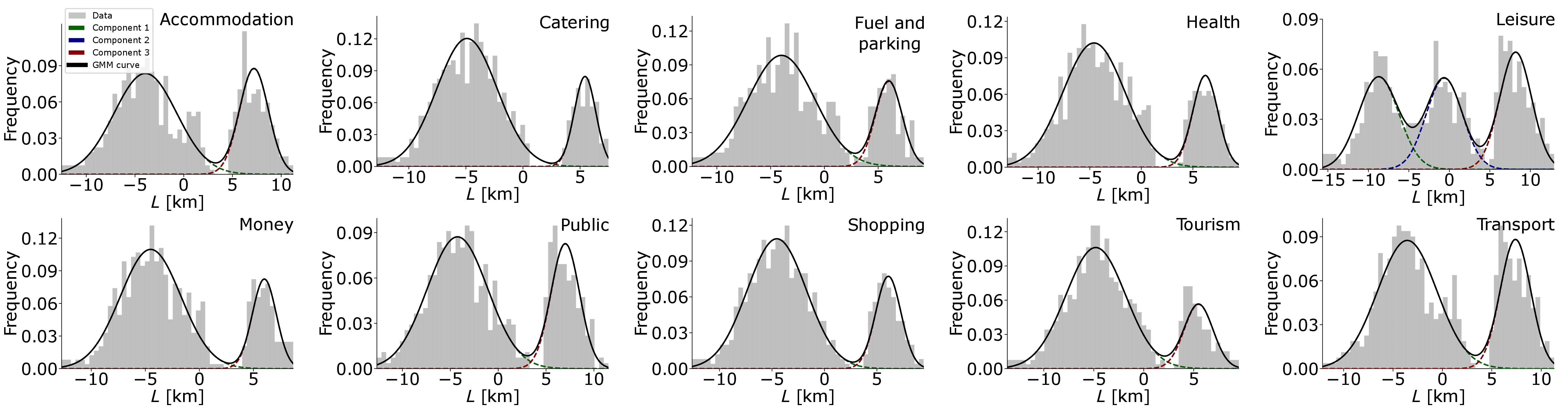}
    \caption{Histograms of $L$ values of all categories for Vienna.}
\end{figure}

\section{Clustering maps}

This section presents all the outcomes of our method (with the Gaussian Mixture Model) used to identify main core (spatial cluster with positive values of $L$ and with the largest mean of the correspondent Gaussian component), periphery (all cells with negative values of $L$ belonging to the Gaussian component with the lowest mean) and intermediate areas (spatial cluster characterized by $L$ values that have the highest probability of belonging to a third Gaussian component whose mean lies between that of the peripheral and the main core region). We can see that all the cities present a main core and a periphery, while some cities show also the presence of intermediate areas.

\begin{figure}[H]
    \centering
\includegraphics[width=\textwidth]{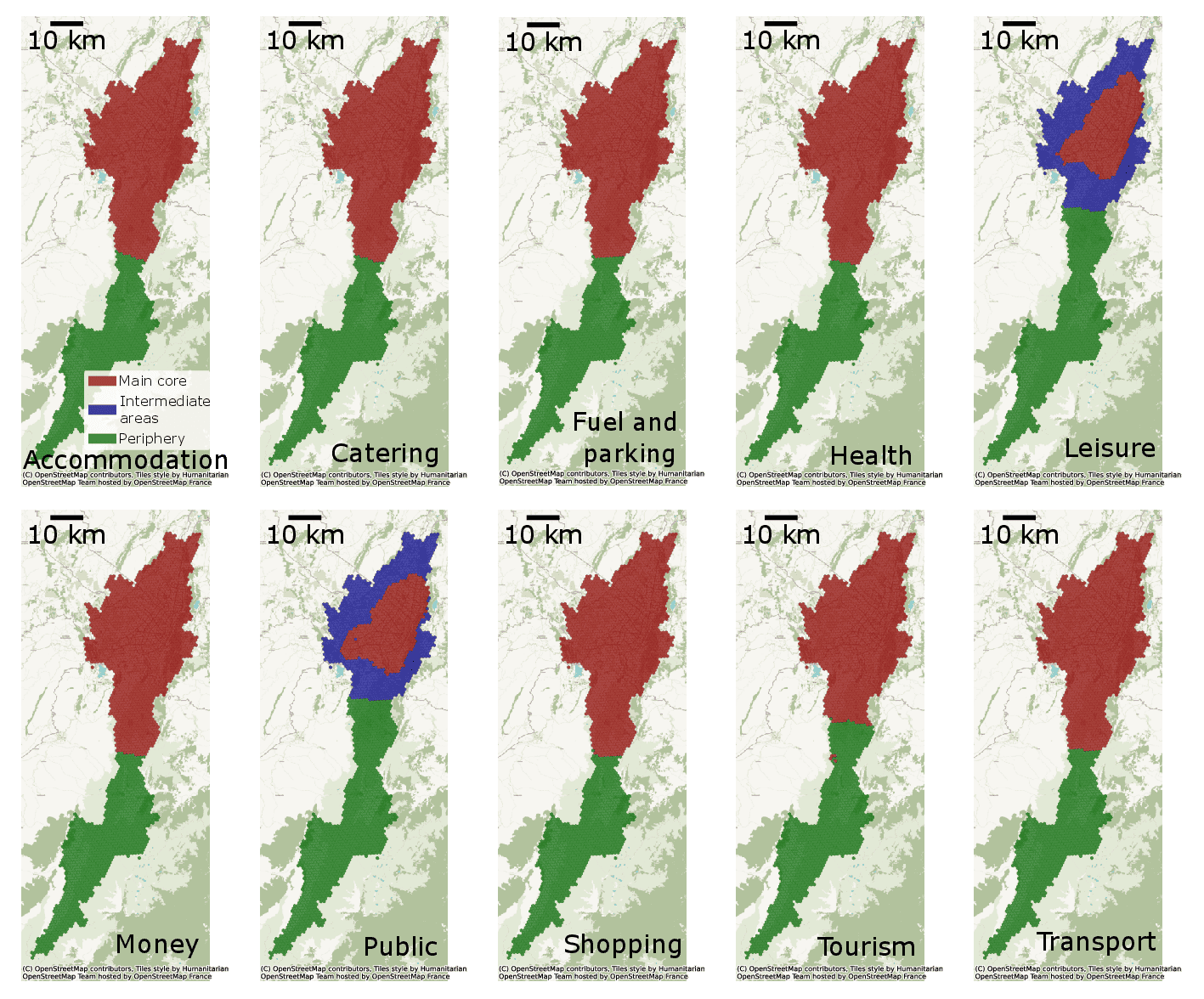}
    \caption{Maps of Bogota with the hexagons colored according to the outcome of the method adopted to identify the components of the urban structure.}
\end{figure}

\begin{figure}[H]
    \centering
\includegraphics[width=\textwidth]{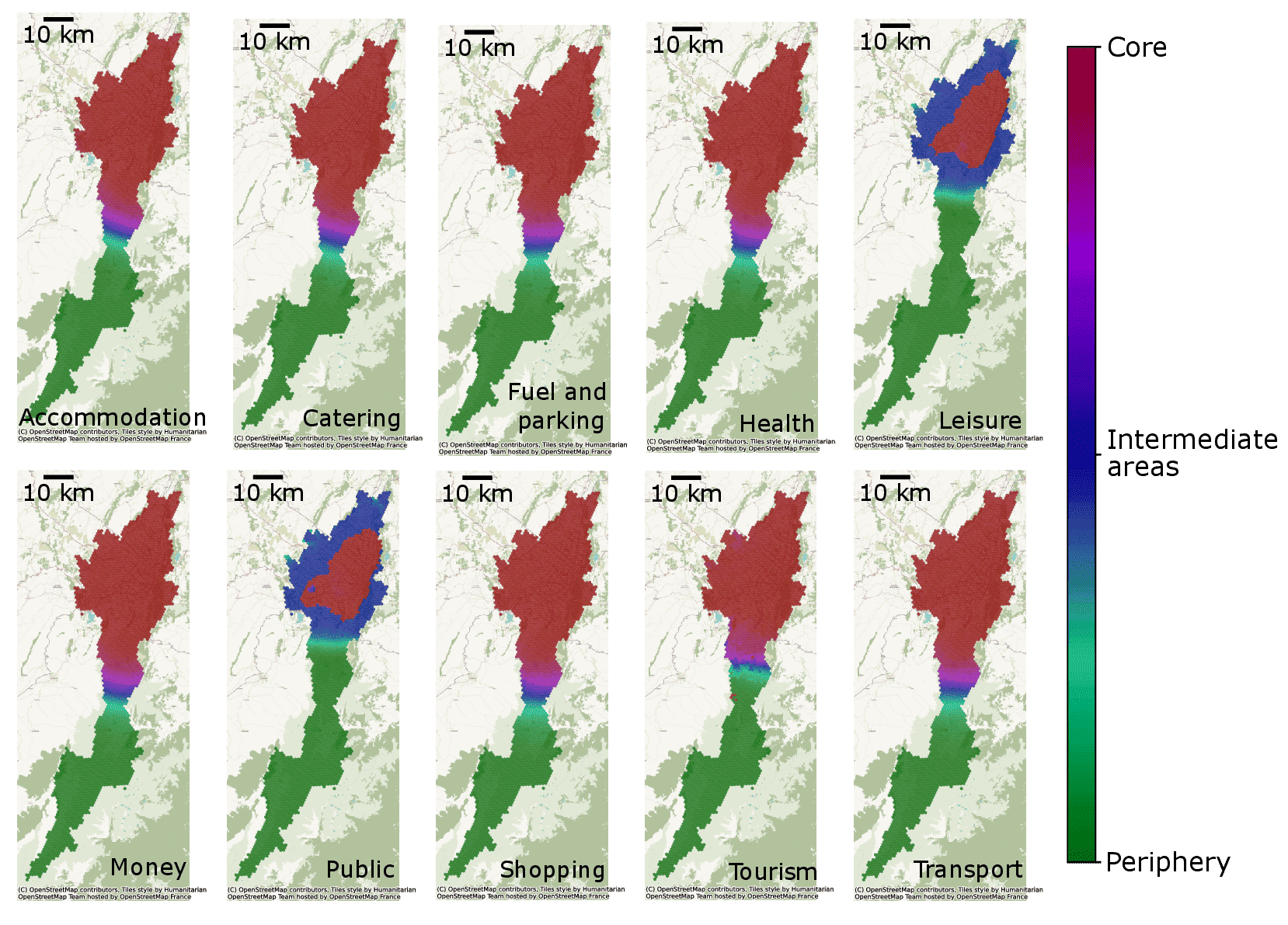}
    \caption{Maps of Bogota that show the spatial distribution of urban zones, using a color gradient, which ranging from green to red, represents the likelihood of areas being classified as part of the periphery, intermediate areas or the main core.}
\end{figure}

\begin{figure}[H]
    \centering
\includegraphics[width=\textwidth]{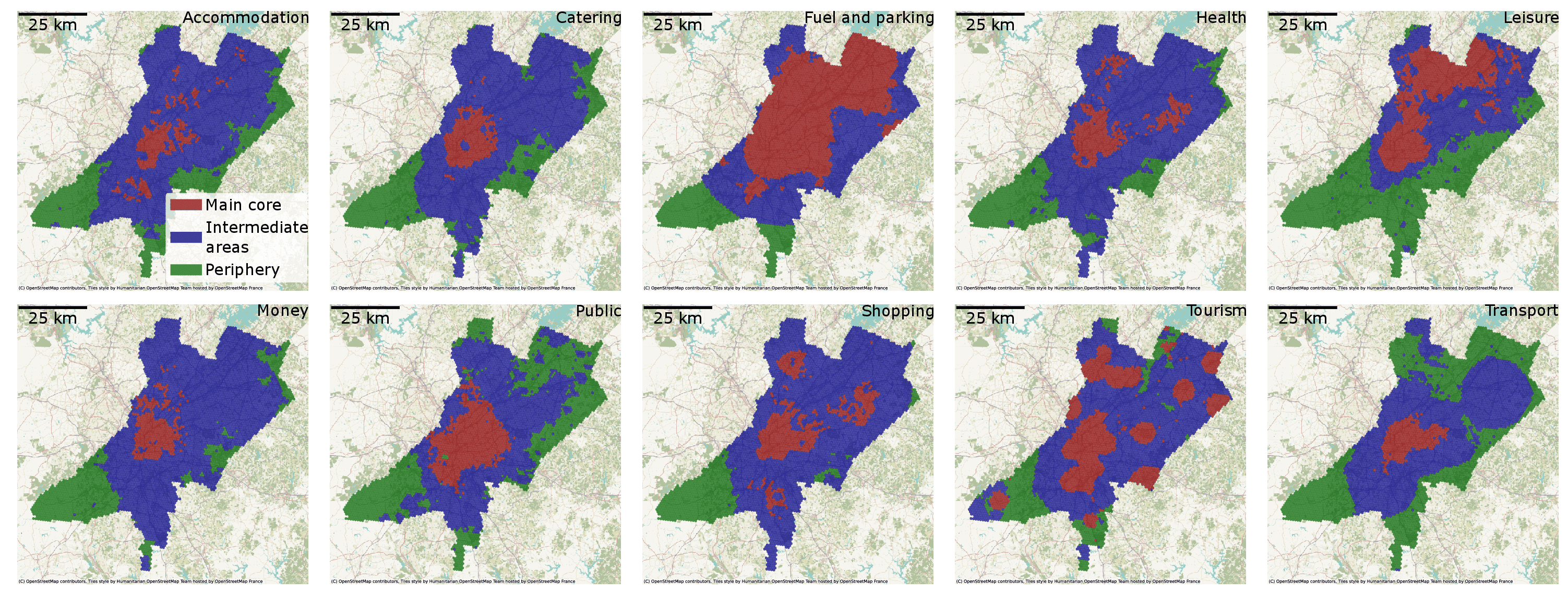}
    \caption{Maps of Atlanta with the hexagons colored according to the outcome of the method adopted to identify the components of the urban structure.}
\end{figure}

\begin{figure}[H]
    \centering
\includegraphics[width=\textwidth]{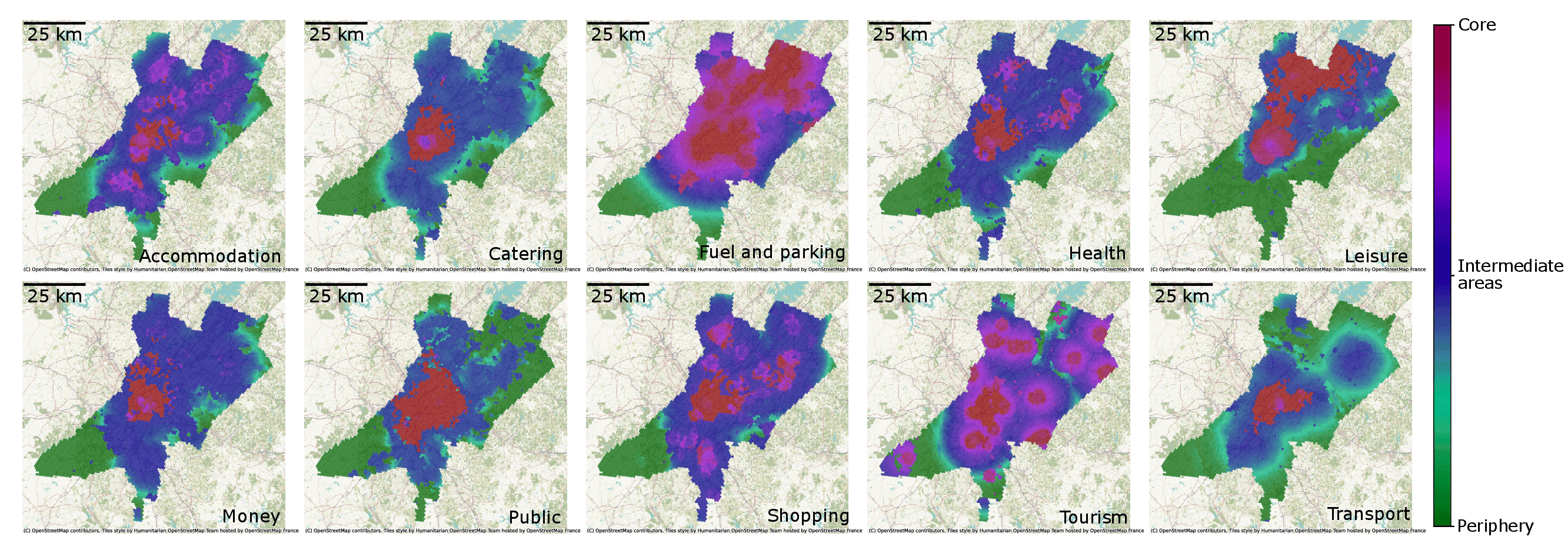}
    \caption{Maps of Atlanta that show the spatial distribution of urban zones, using a color gradient, which ranging from green to red, represents the likelihood of areas being classified as part of the periphery, intermediate areas or the main core.}
\end{figure}

\begin{figure}[H]
    \centering
\includegraphics[width=\textwidth]{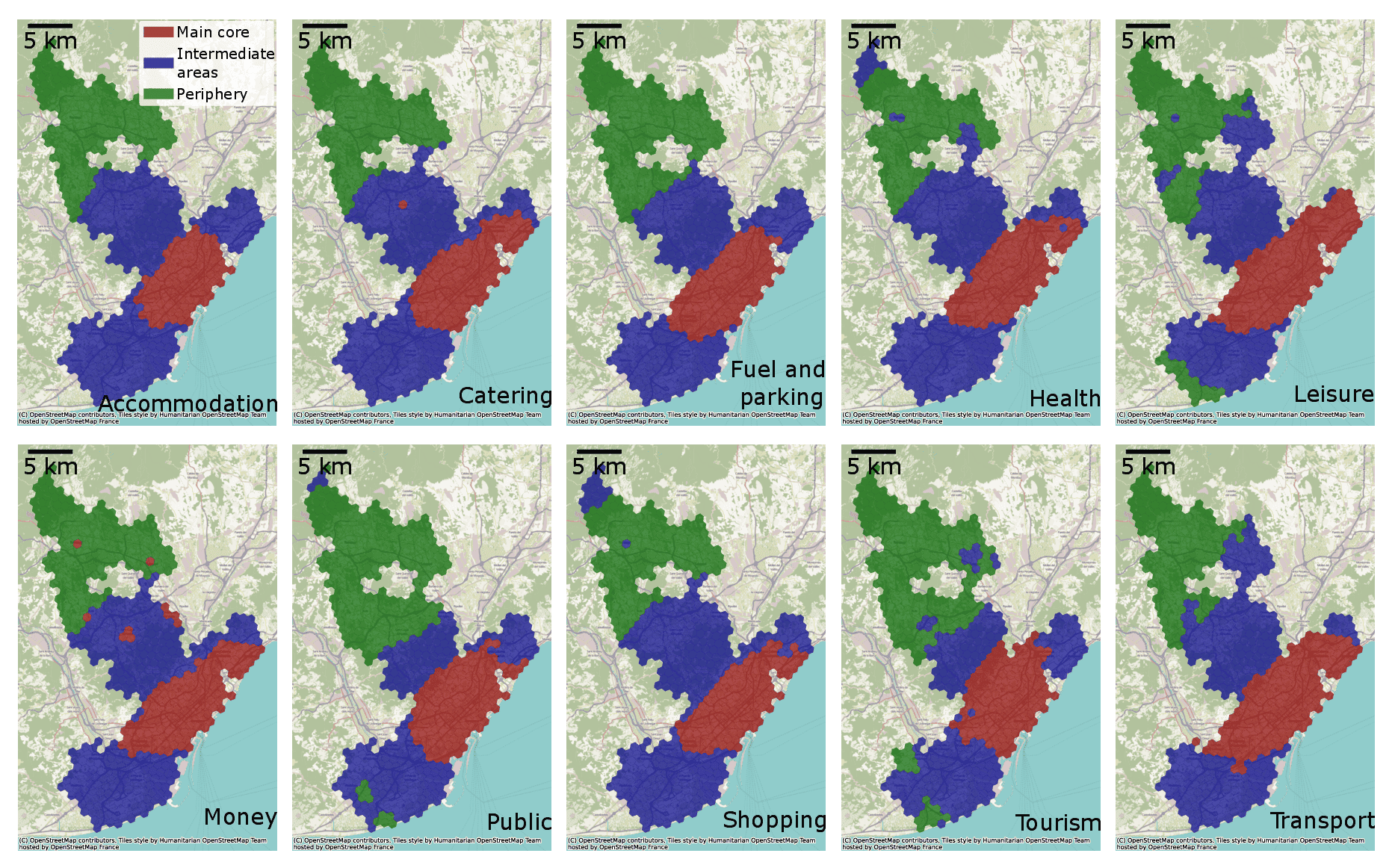}
    \caption{Maps of Barcelona with the hexagons colored according to the outcome of the method adopted to identify the components of the urban structure.}
\end{figure}

\begin{figure}[H]
    \centering
\includegraphics[width=\textwidth]{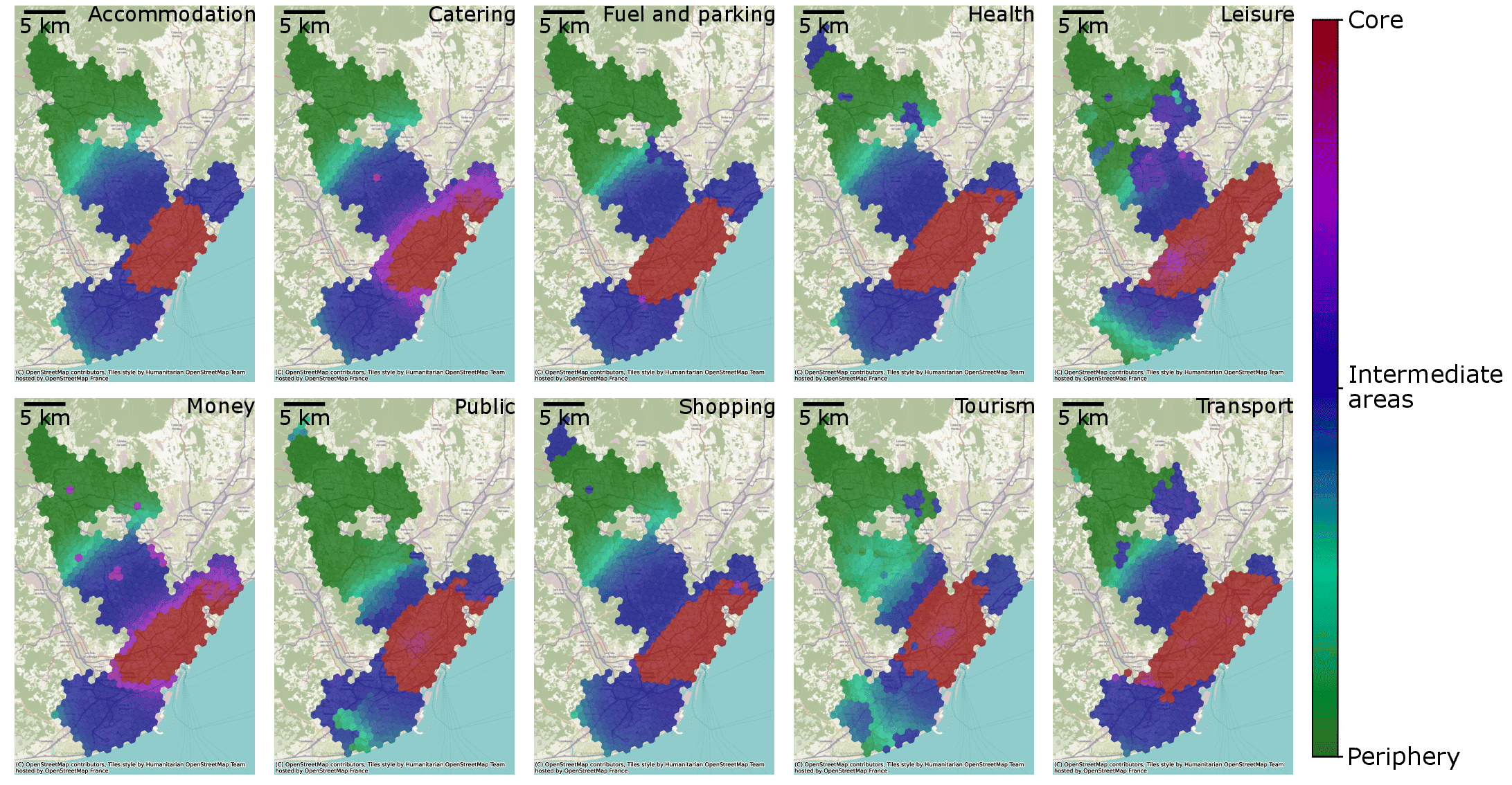}
    \caption{Maps of Barcelona that show the spatial distribution of urban zones, using a color gradient, which ranging from green to red, represents the likelihood of areas being classified as part of the periphery, intermediate areas or the main core.}
\end{figure}

\begin{figure}[H]
    \centering
\includegraphics[width=\textwidth]{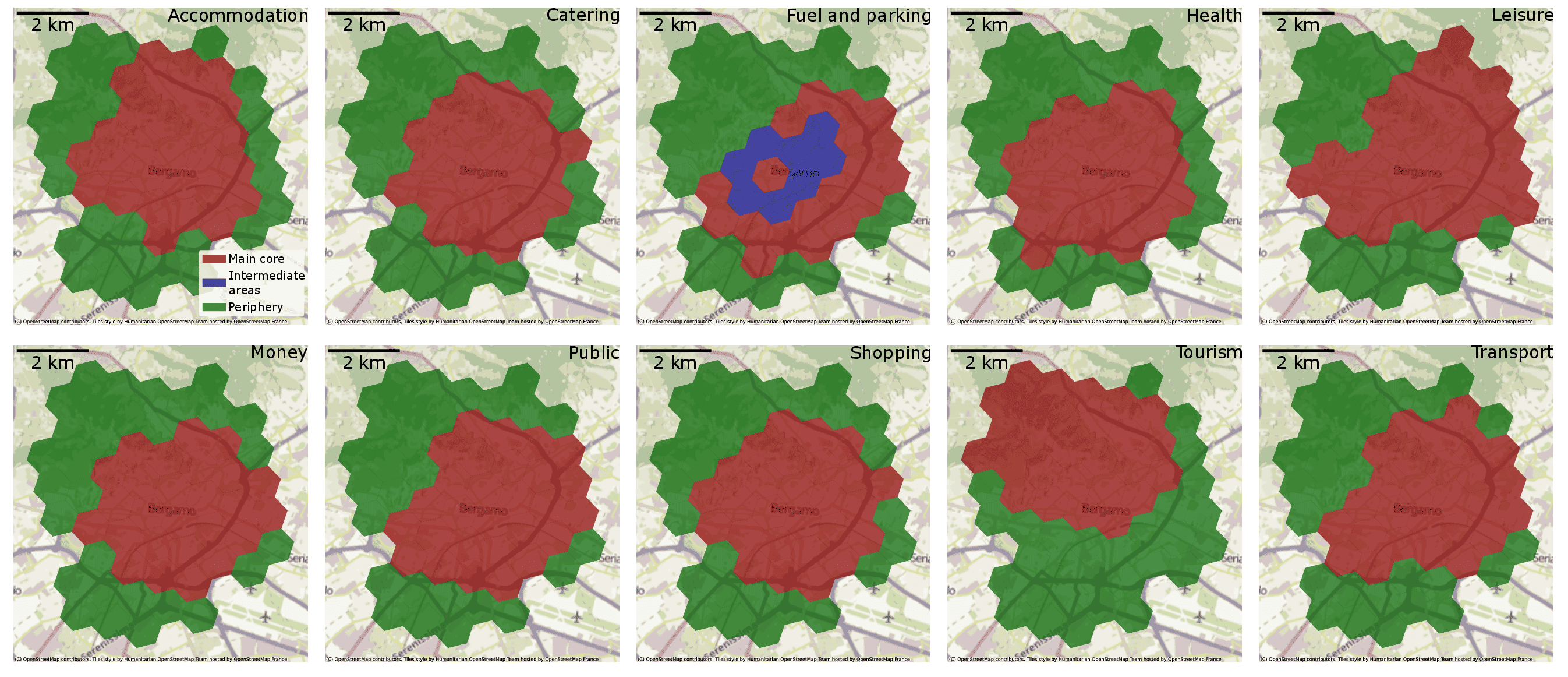}
    \caption{Maps of Bergamo with the hexagons colored according to the outcome of the method adopted to identify the components of the urban structure.}
\end{figure}

\begin{figure}[H]
    \centering
\includegraphics[width=\textwidth]{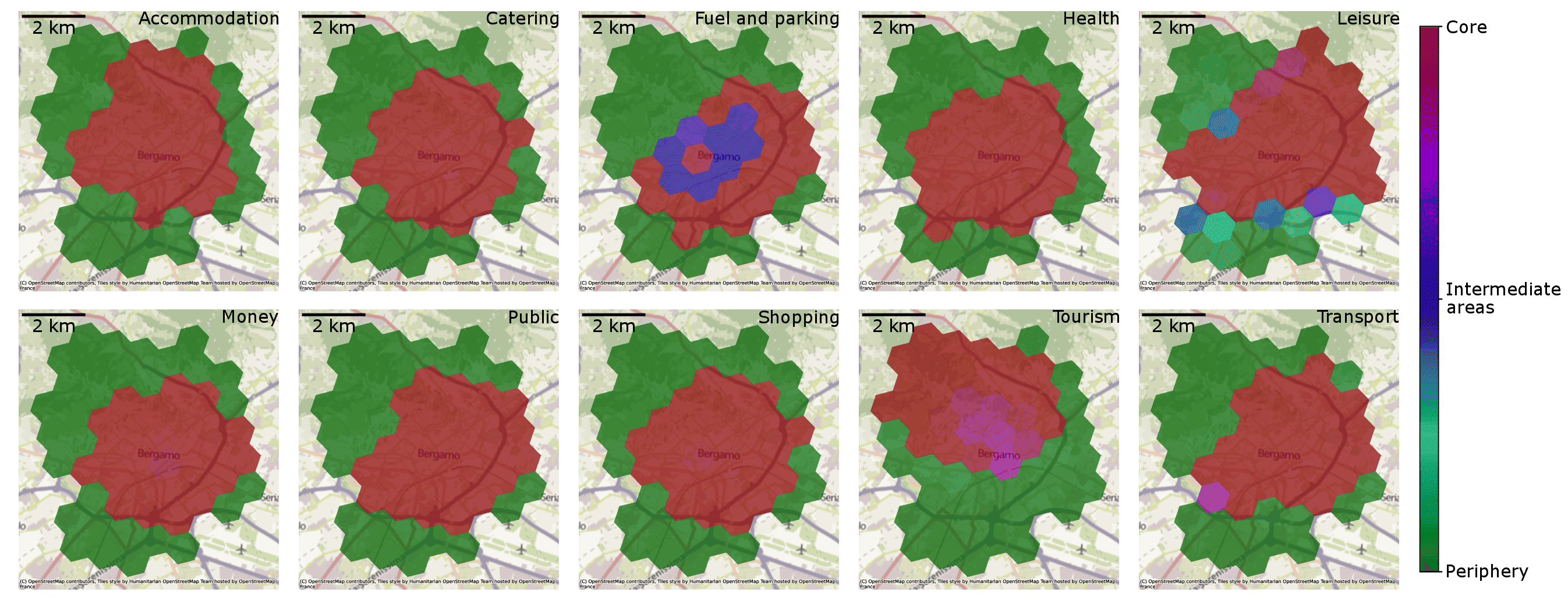}
    \caption{Maps of Bergamo that show the spatial distribution of urban zones, using a color gradient, which ranging from green to red, represents the likelihood of areas being classified as part of the periphery, intermediate areas or the main core.}
\end{figure}

\begin{figure}[H]
    \centering
\includegraphics[width=\textwidth]{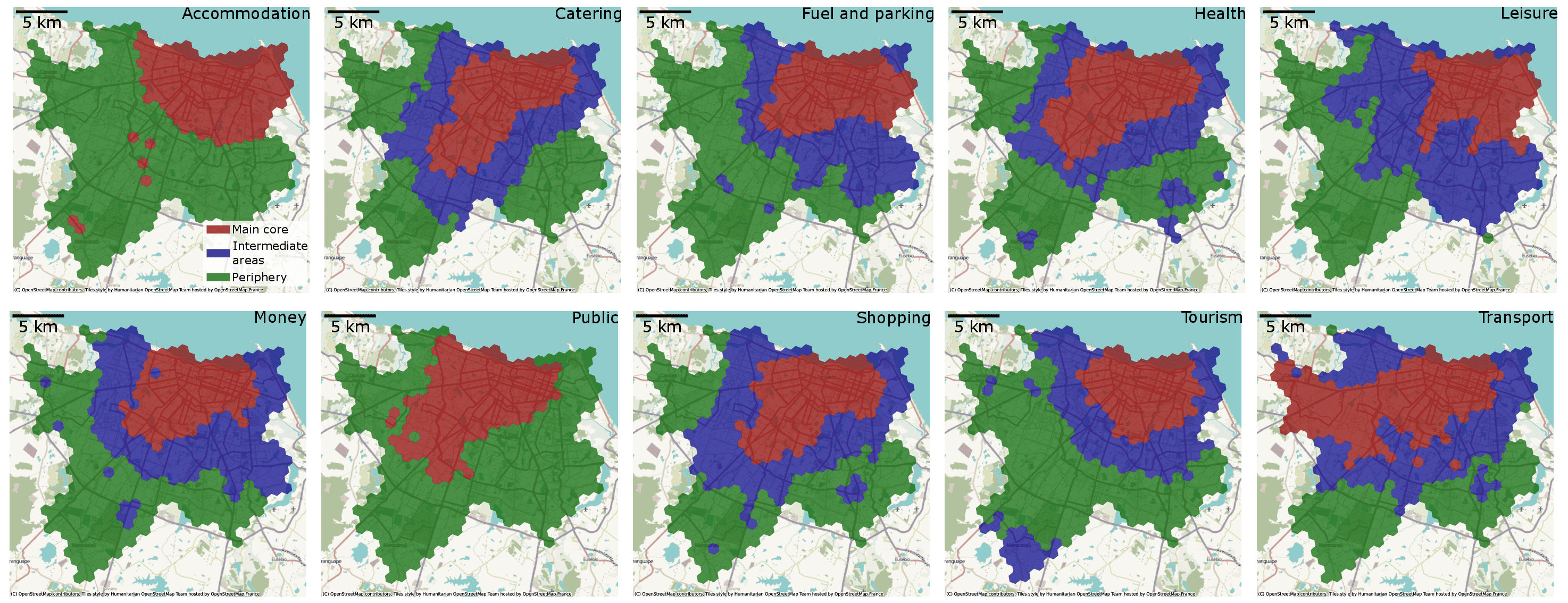}
    \caption{Maps of Fortaleza with the hexagons colored according to the outcome of the method adopted to identify the components of the urban structure.}
\end{figure}

\begin{figure}[H]
    \centering
\includegraphics[width=\textwidth]{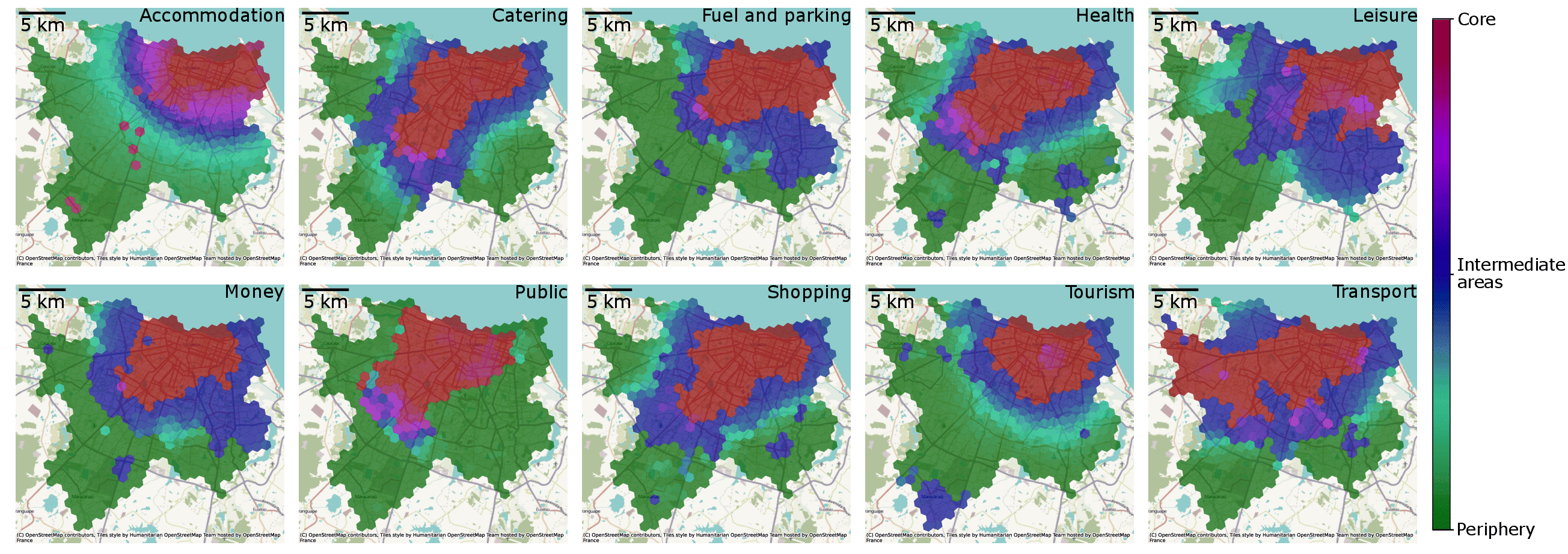}
    \caption{Maps of Fortaleza that show the spatial distribution of urban zones, using a color gradient, which ranging from green to red, represents the likelihood of areas being classified as part of the periphery, intermediate areas or the main core.}
\end{figure}

\begin{figure}[H]
    \centering
\includegraphics[width=\textwidth]{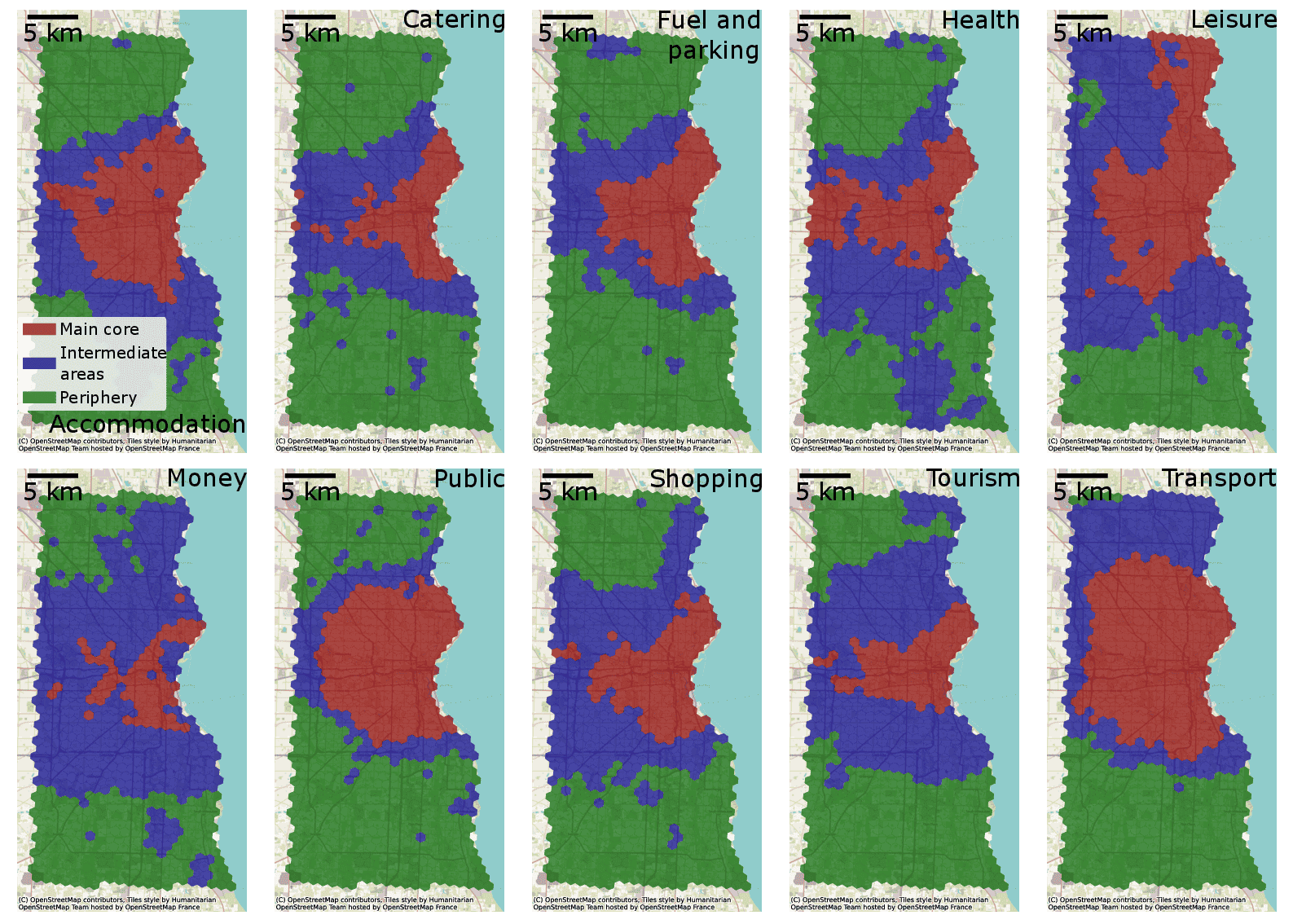}
    \caption{Maps of Milwaukee with the hexagons colored according to the outcome of the method adopted to identify the components of the urban structure.}
\end{figure}

\begin{figure}[H]
    \centering
\includegraphics[width=\textwidth]{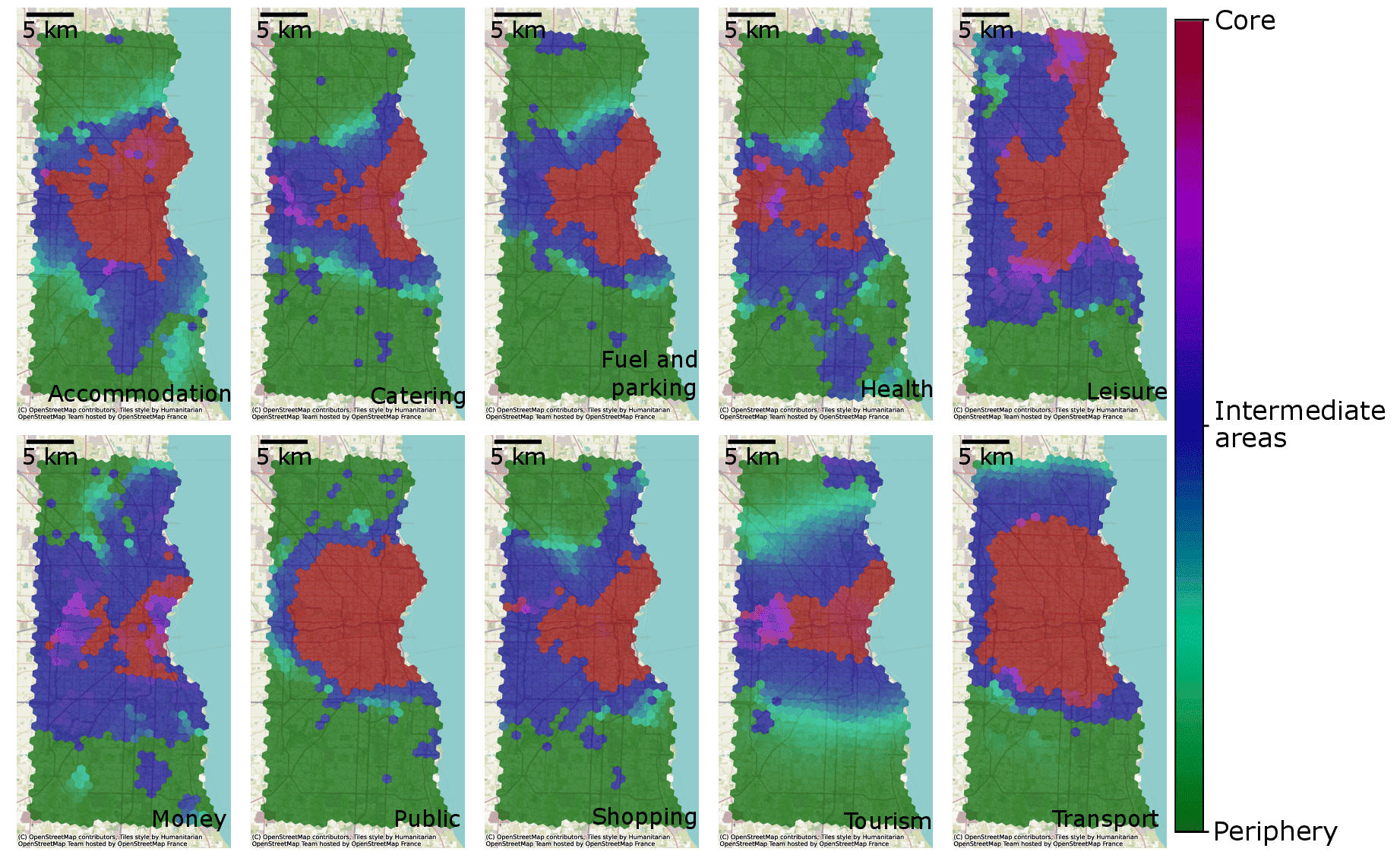}
    \caption{Maps of Milwaukee that show the spatial distribution of urban zones, using a color gradient, which ranging from green to red, represents the likelihood of areas being classified as part of the periphery, intermediate areas or the main core.}
\end{figure}

\begin{figure}[H]
    \centering
\includegraphics[width=\textwidth]{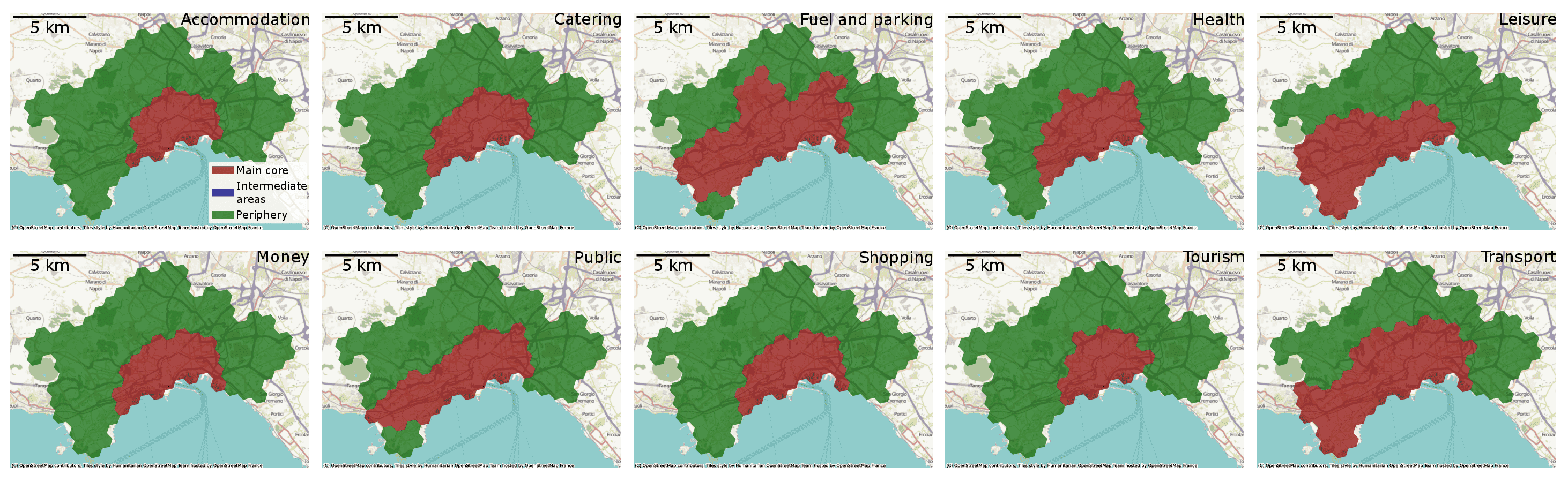}
    \caption{Maps of Naples with the hexagons colored according to the outcome of the method adopted to identify the components of the urban structure.}
\end{figure}

\begin{figure}[H]
    \centering
\includegraphics[width=\textwidth]{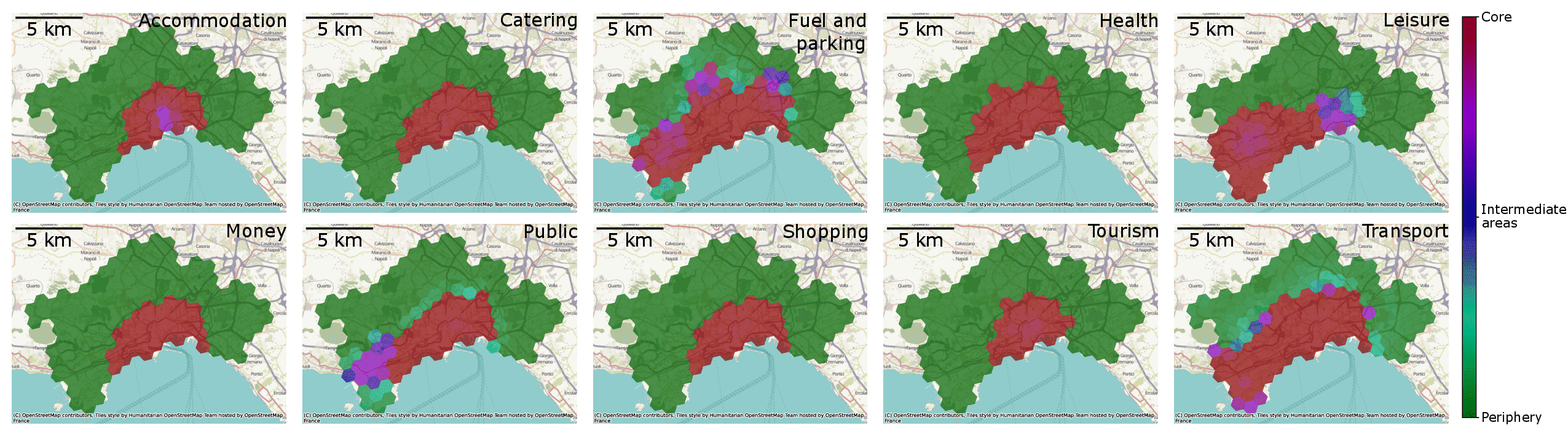}
    \caption{Maps of Naples that show the spatial distribution of urban zones, using a color gradient, which ranging from green to red, represents the likelihood of areas being classified as part of the periphery, intermediate areas or the main core.}
\end{figure}

\begin{figure}[H]
    \centering
\includegraphics[width=\textwidth]{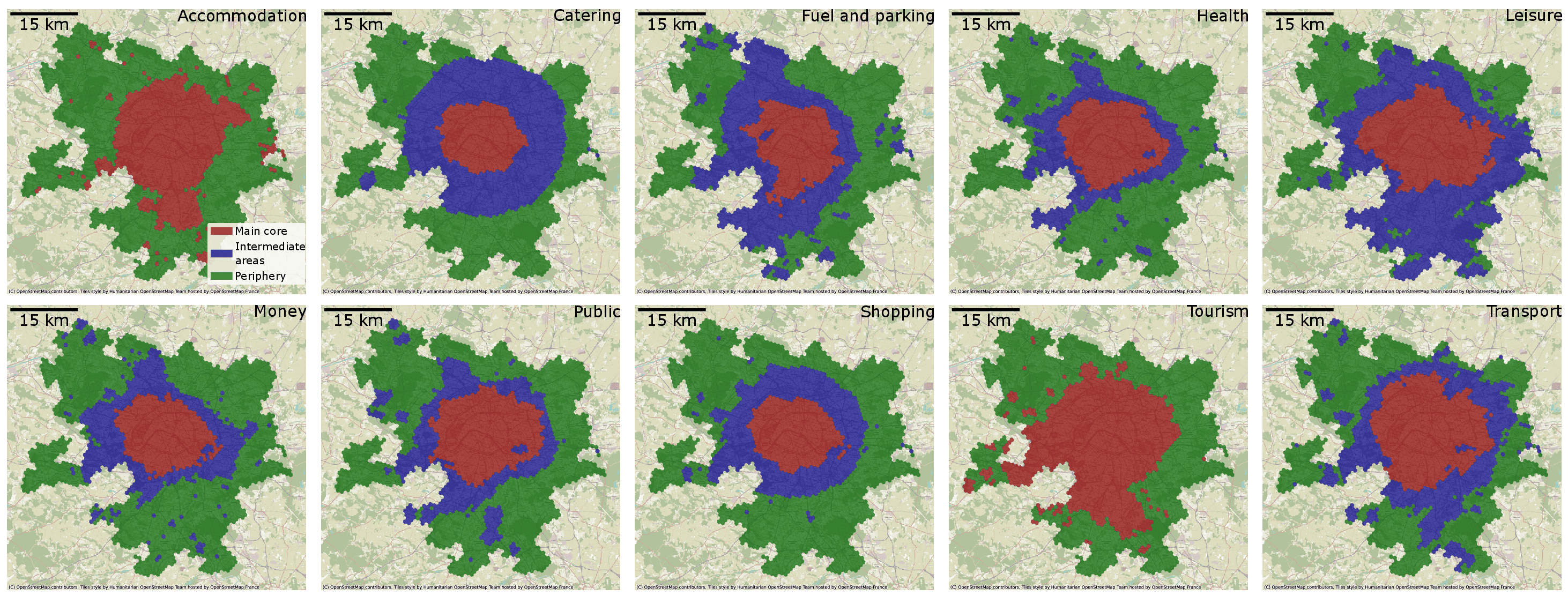}
    \caption{Maps of Paris with the hexagons colored according to the outcome of the method adopted to identify the components of the urban structure.}
\end{figure}

\begin{figure}[H]
    \centering
\includegraphics[width=\textwidth]{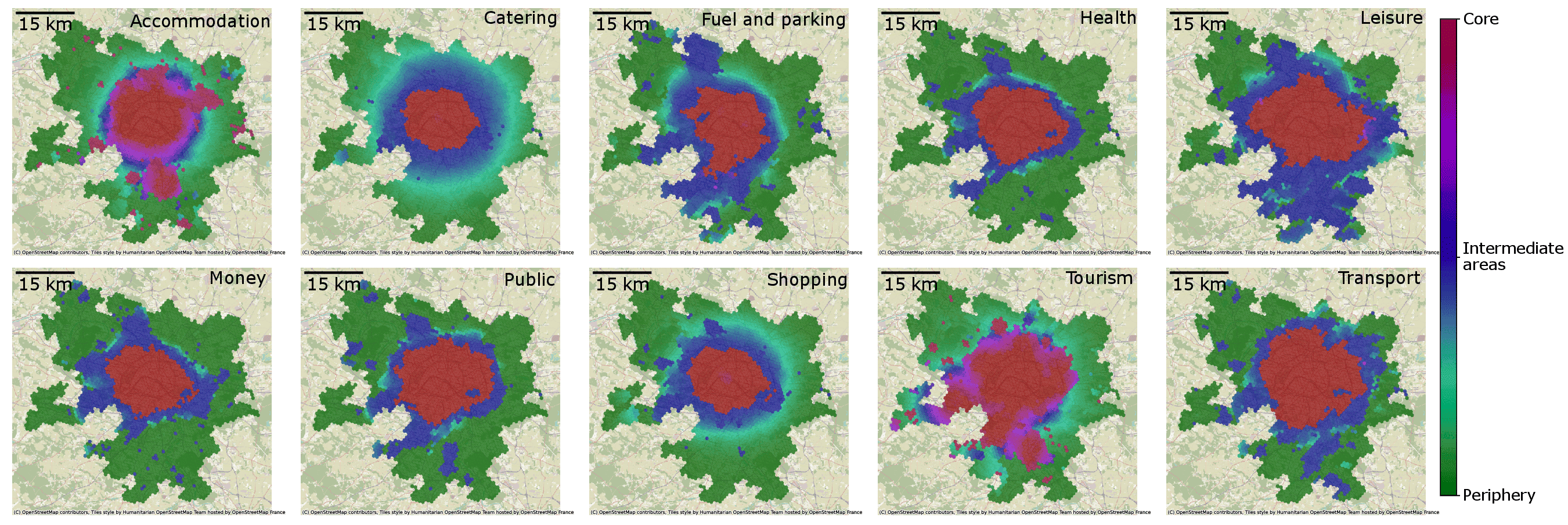}
    \caption{Maps of Paris that show the spatial distribution of urban zones, using a color gradient, which ranging from green to red, represents the likelihood of areas being classified as part of the periphery, intermediate areas or the main core.}
\end{figure}

\begin{figure}[H]
    \centering
\includegraphics[width=\textwidth]{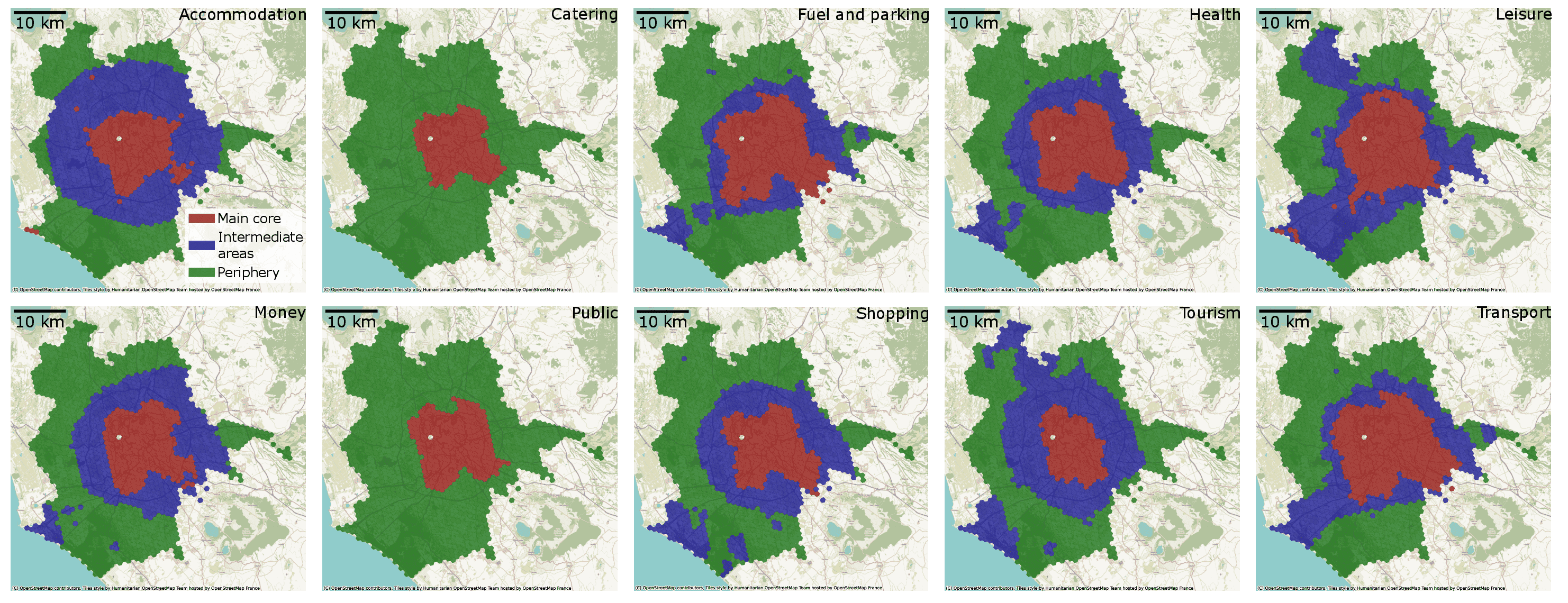}
    \caption{Maps of Rome with the hexagons colored according to the outcome of the method adopted to identify the components of the urban structure.}
\end{figure}

\begin{figure}[H]
    \centering
\includegraphics[width=\textwidth]{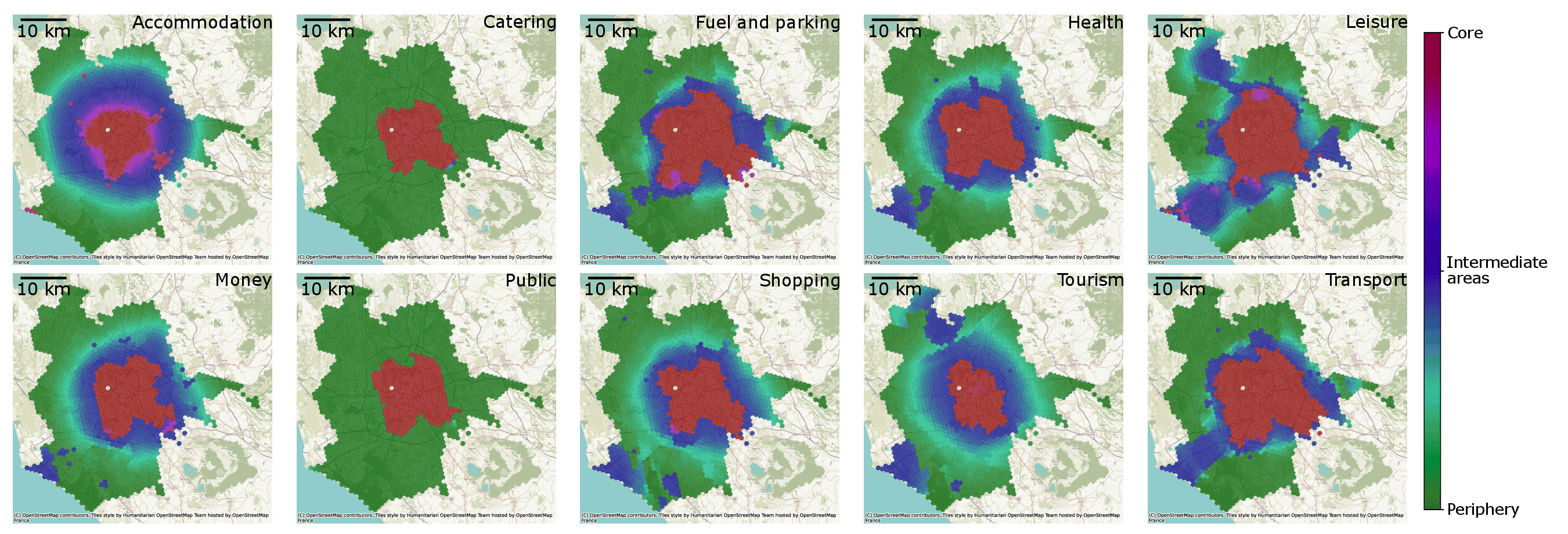}
    \caption{Maps of Rome that show the spatial distribution of urban zones, using a color gradient, which ranging from green to red, represents the likelihood of areas being classified as part of the periphery, intermediate areas or the main core.}
\end{figure}

\begin{figure}[H]
    \centering
\includegraphics[width=\textwidth]{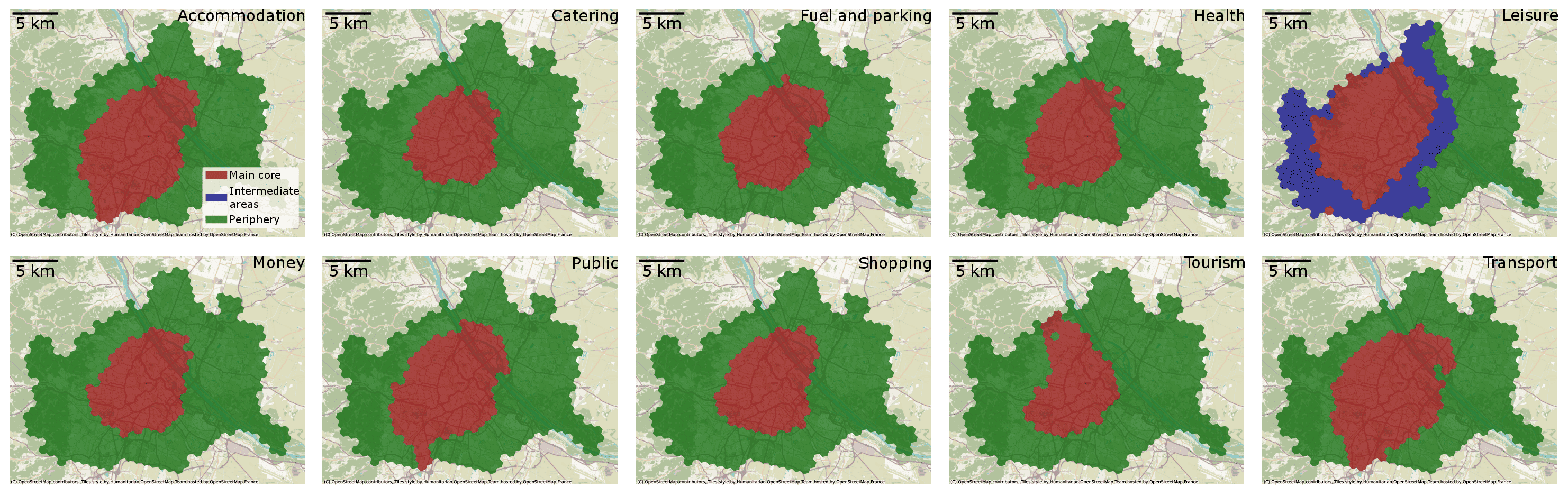}
    \caption{Maps of Vienna with the hexagons colored according to the outcome of the method adopted to identify the components of the urban structure.}
\end{figure}

\begin{figure}[H]
    \centering
\includegraphics[width=\textwidth]{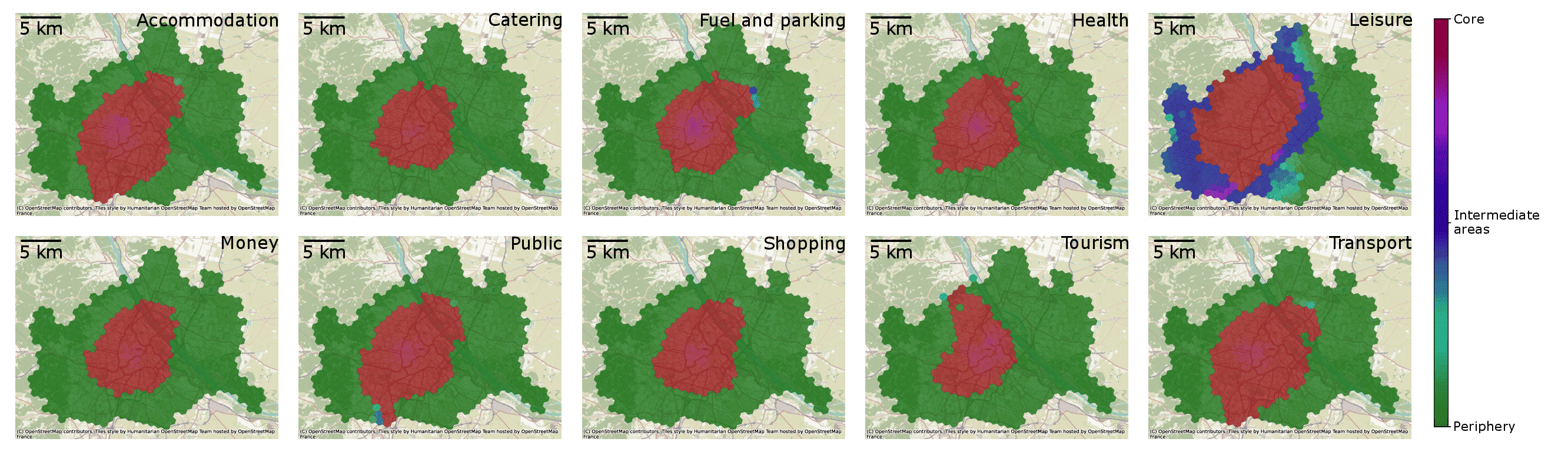}
    \caption{Maps of Vienna that show the spatial distribution of urban zones, using a color gradient, which ranging from green to red, represents the likelihood of areas being classified as part of the periphery, intermediate areas or the main core.}
\end{figure}

\section{Distributions of $L$ weighed with population}

We weighted the histogram of $L$ values with the population living in each hexagon. We observed that peaks are always present, but now their height is different. If we think to our interpretation of the peaks through our method to identify the components of the city, we observed that the peaks whose height becomes smaller are the ones related to the peripheral area: the hexagons characterized by negative values of $L$ show a low number of people living there. So we have areas with a bad accessibility, but the fraction of population living there is low. The peaks related to the functional core and the secondary cores are the most populated parts of the city.

\begin{figure}[H]
    \centering
\includegraphics[width=\textwidth]{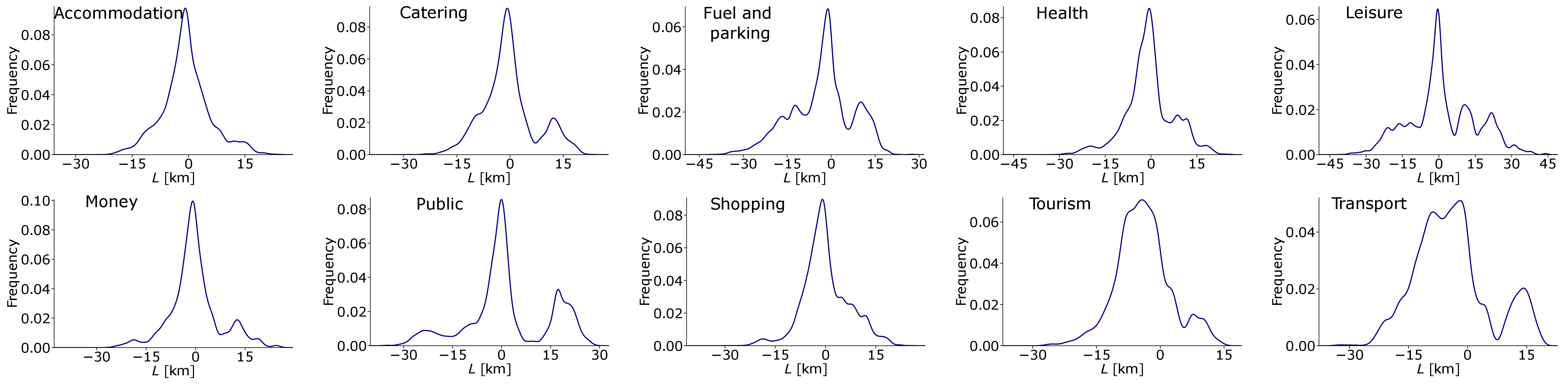}
    \caption{Histograms of $L$ values weighed with population for all categories of Atlanta.}
\end{figure}

\begin{figure}[H]
    \centering
\includegraphics[width=\textwidth]{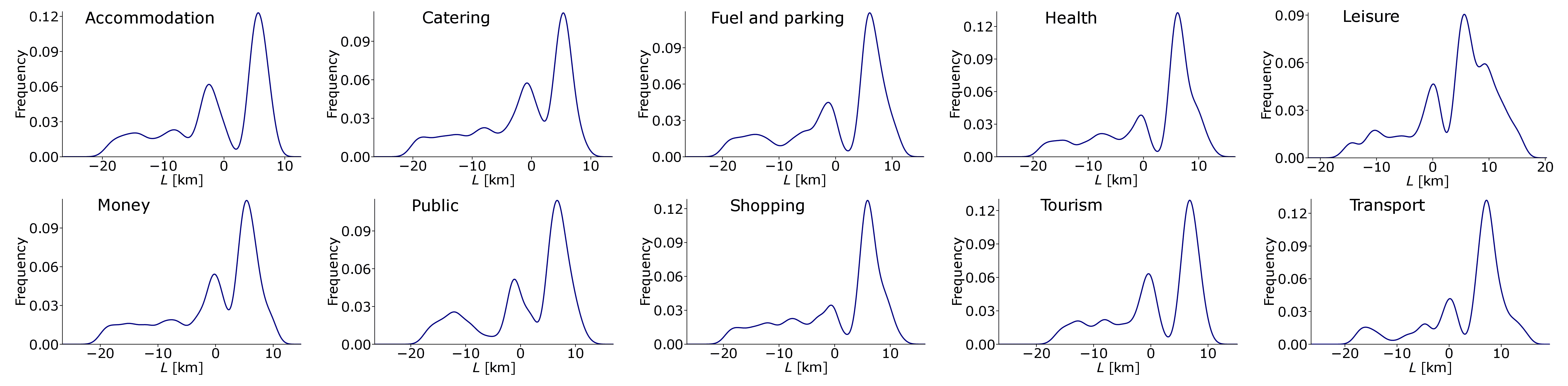}
    \caption{Histograms of $L$ values weighed with population for all categories of Barcelona.}
\end{figure}

 \begin{figure}[H]
    \centering
\includegraphics[width=\textwidth]{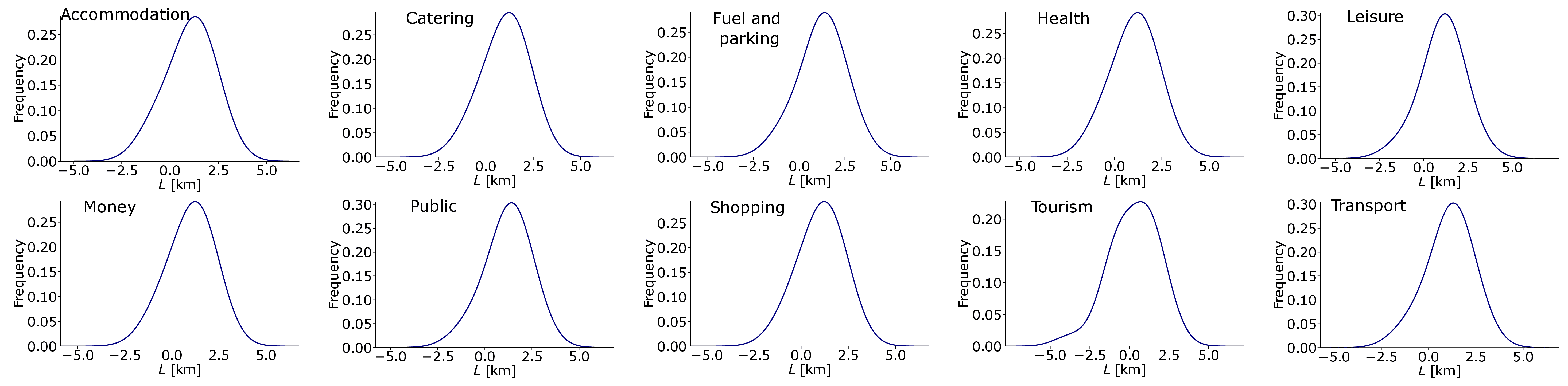}
    \caption{Histograms of $L$ values weighed with population for all categories of Bergamo.}
\end{figure}

\begin{figure}[H]
    \centering
\includegraphics[width=\textwidth]{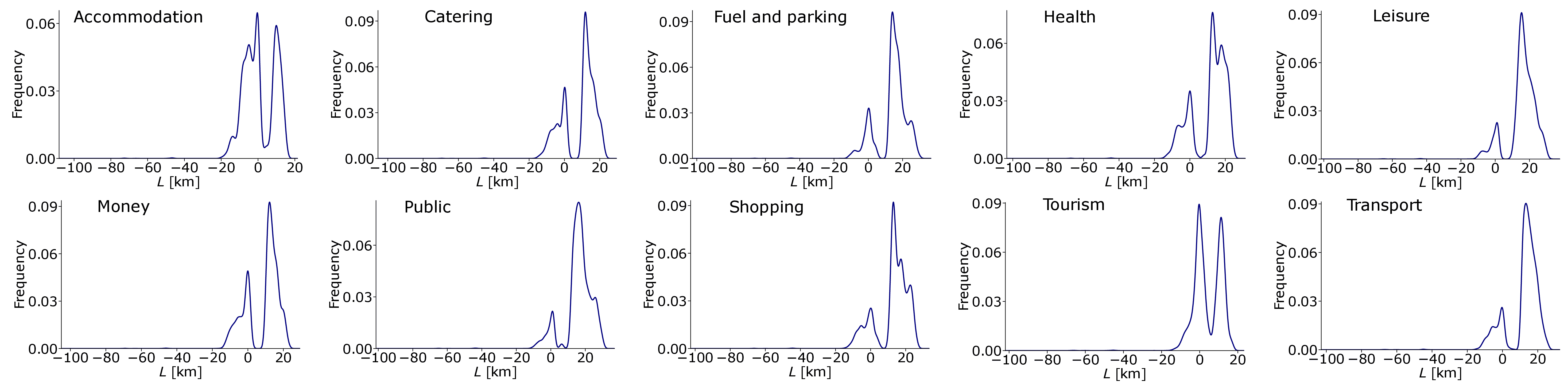}
    \caption{Histograms of $L$ values weighed with population for all categories of Bogota.}
\end{figure}

 \begin{figure}[H]
    \centering
\includegraphics[width=\textwidth]{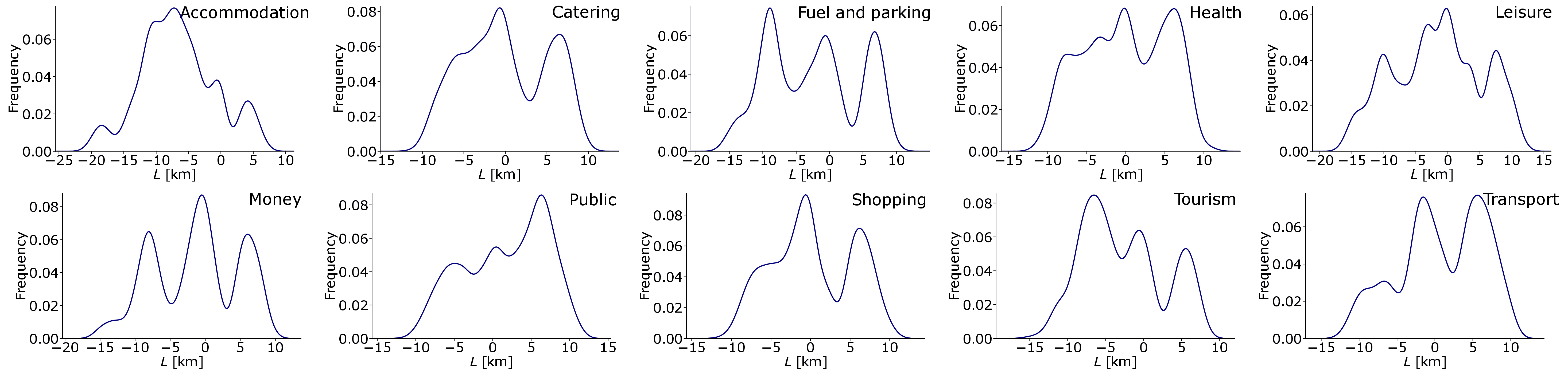}
    \caption{Histograms of $L$ values weighed with population for all categories of Fortaleza.}
\end{figure}

\begin{figure}[H]
    \centering
\includegraphics[width=\textwidth]{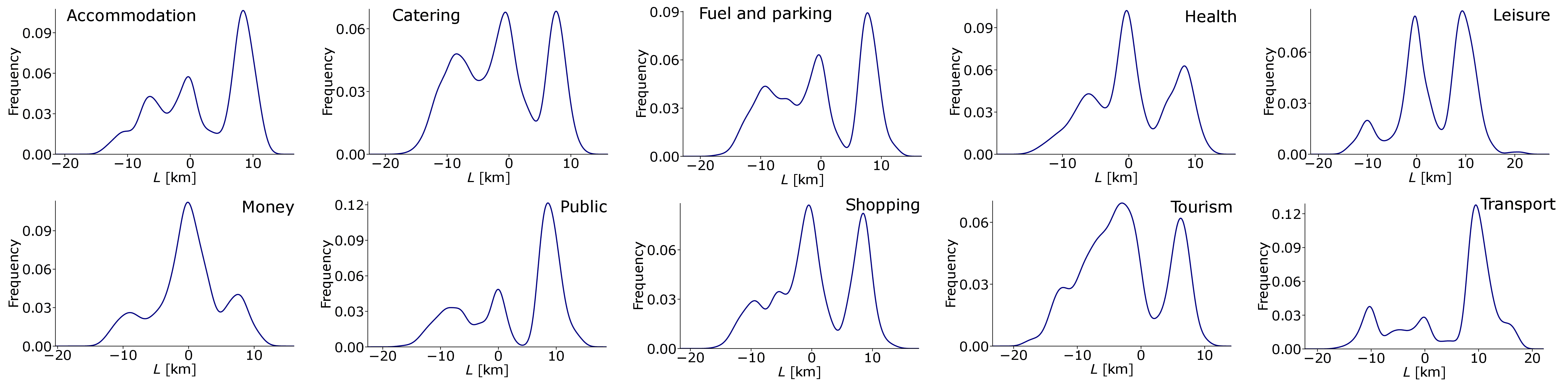}
    \caption{Histograms of $L$ values weighed with population for all categories  of Milwaukee.}
\end{figure}

\begin{figure}[H]
    \centering
\includegraphics[width=\textwidth]{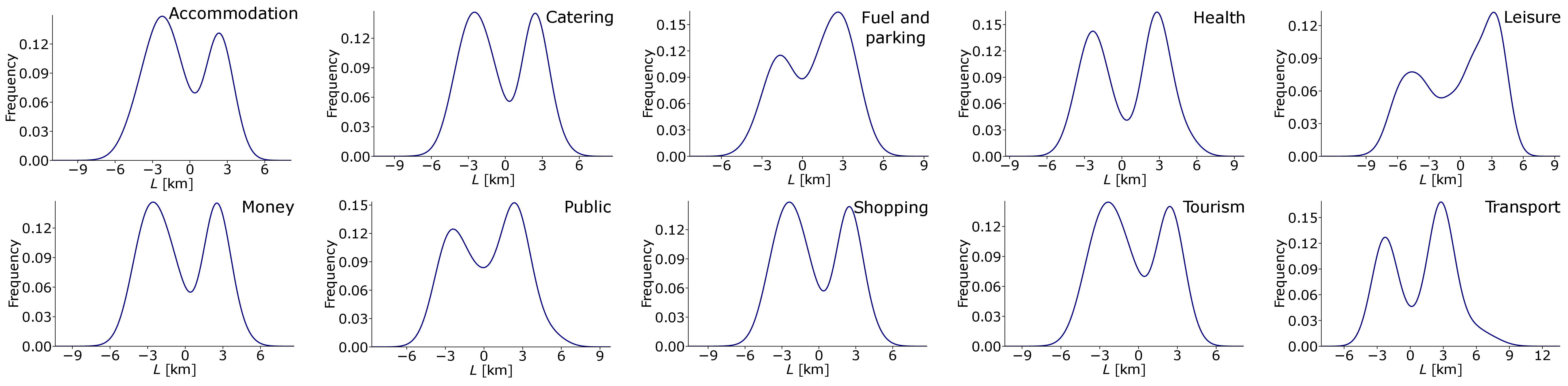}
    \caption{Histograms of $L$ values weighed with population for all categories of Naples.}
\end{figure}

 \begin{figure}[H]
    \centering
\includegraphics[width=\textwidth]{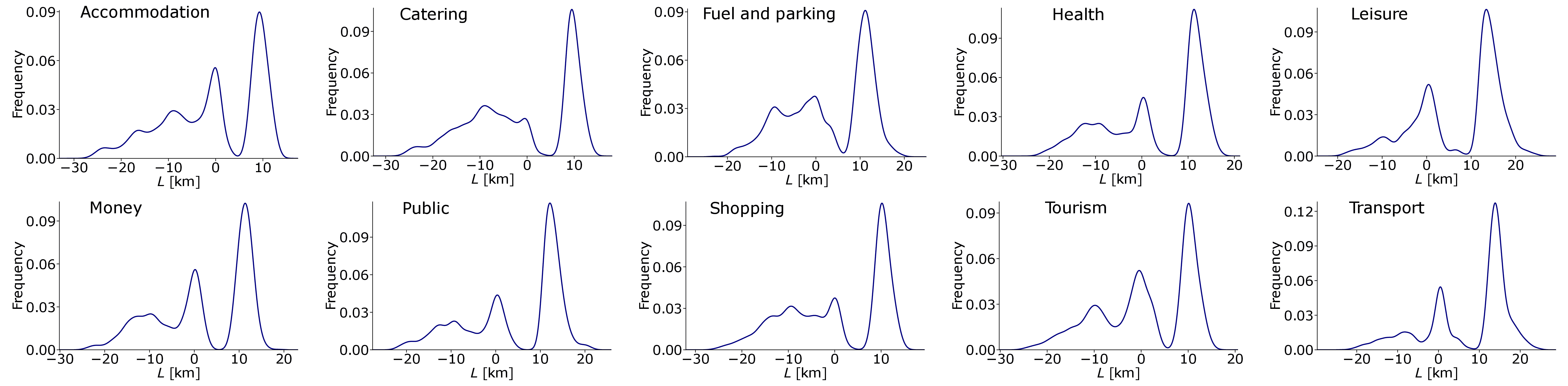}
    \caption{Histograms of $L$ values weighed with population for all categories of Paris.}
\end{figure}

\begin{figure}[H]
    \centering
\includegraphics[width=\textwidth]{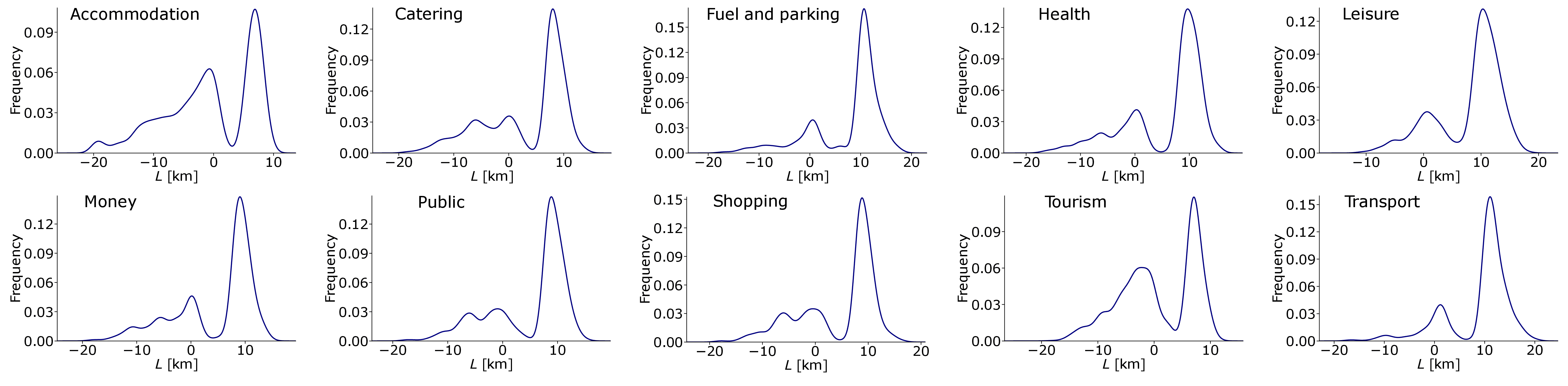}
    \caption{Histograms of $L$ values weighed with population for all categories of Rome.}
\end{figure}

 \begin{figure}[H]
    \centering
\includegraphics[width=\textwidth]{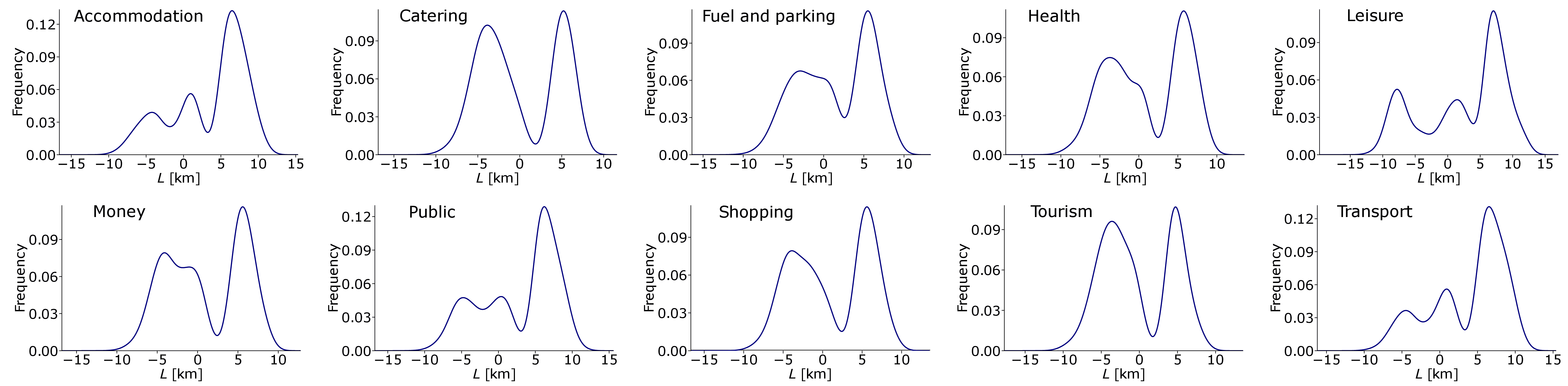}
    \caption{Histograms of $L$ values weighed with population for all categories of Vienna.}
\end{figure}